\documentclass[a4paper,preprint,number,sort&compress]{elsarticle}
\usepackage[utf8]{inputenc}

\makeatletter
\def\ps@pprintTitle{%
 \let\@oddhead\@empty
 \let\@evenhead\@empty
 \def\@oddfoot{\centerline{\thepage}}%
 \let\@evenfoot\@oddfoot}
\makeatother

\usepackage{placeins} 
\usepackage{subcaption} 
\usepackage{mathtools} 
\usepackage{amssymb} 
\usepackage{multirow} 
\usepackage[ruled,vlined,linesnumbered]{algorithm2e} 
\usepackage[short]{optidef} 
\usepackage{etoolbox} 
\usepackage{textcomp} 
\usepackage{xparse} 
\usepackage[hyphens]{url} 
\usepackage{cleveref} 

\robustify{\label}

\graphicspath{{./img/}}

\allowdisplaybreaks

\begin{document}
\newcommand{\var}{\text{var}}
\newcommand{\Rz}{\mathcal{R}_{c}}
\newcommand{\tabbox}[2]{\vtop{\hbox{\strut #1}\hbox{\strut #2}}}

\begin{frontmatter}
    \title{Likelihood-based estimation and prediction for a measles outbreak in Samoa}

    \author[1]{David Wu \corref{cor}} 
    \ead{dwu402@aucklanduni.ac.nz}
    \author[2]{Helen Petousis-Harris}
    \author[2]{Janine Paynter}
    \author[1,3]{Vinod Suresh}
    \author[1]{Oliver J. Maclaren}
    
    \affiliation[1]{
        organization={Department of Engineering Science, University of Auckland},
        addressline={Grafton},
        city={Auckland},
        postcode={1010},
        country={New Zealand},
    }
    \affiliation[2]{
        organization={Department of General Practice and Primary Health Care, University of Auckland},
        addressline={Grafton},
        city={Auckland},
        postcode={1023},
        country={New Zealand},
    }
    \affiliation[3]{
        organization={Auckland Bioengineering Institute, University of Auckland},
        addressline={Grafton},
        city={Auckland},
        postcode={1010},
        country={New Zealand}
    }
    
    \cortext[cor]{Corresponding author}

\begin{abstract}
    Prediction of the progression of an infectious disease outbreak is important for planning and coordinating a response. Differential equations are often used to model an epidemic outbreak's behaviour but are challenging to parameterise. Furthermore, these models can suffer from misspecification, which biases predictions and parameter estimates. Stochastic models can help with misspecification but are even more expensive to simulate and perform inference with. Here, we develop an explicitly likelihood-based variation of the generalised profiling method as a tool for prediction and inference under model misspecification. Our approach allows us to carry out identifiability analysis and uncertainty quantification using profile likelihood-based methods without the need for marginalisation.
    We provide justification for this approach by introducing a new interpretation of the model approximation component as a stochastic constraint. This preserves the rationale for using profiling rather than integration to remove nuisance parameters while also providing a link back to stochastic models. We applied an initial version of this method during an outbreak of measles in Samoa in 2019-2020 and found that it achieved relatively fast, accurate predictions. Here we present the most recent version of our method and its application to this measles outbreak, along with additional validation.
\end{abstract}


\end{frontmatter}

\section{Introduction}
The prediction of the progression of epidemic outbreaks of infectious diseases is notoriously difficult, with multiple confounding sources of uncertainty. Despite this, even approximate predictions are important in informing the response to outbreaks. This importance is emphasised in areas where resources restrict response infrastructure or rapid response is required. Early predictions are typically model-based, as the available data alone are usually insufficient for understanding the key dynamics of an outbreak.

The models used to characterise the spread of infectious diseases are usually mechanistic models, and these are well-established in the mathematical epidemiological community \cite{anderson92}. From the early deterministic models of \citet{ross1911} and \citet{kermack1927}, to recent developments in both deterministic \cite{schenzle1984,hooker2011,xia2004} and stochastic models \cite{allen2008,allen2011,ferguson2020}, there are a range of models that can be used to describe various aspects of the dynamics of infectious disease outbreaks. However, there are still difficult open problems on how best to relate these models to data and extract meaningful information about model behaviour specific to particular outbreaks \citep{wilke_bergstrom_2020, roda2020}.

More widely, there are many methods for performing parameter inference for mechanistic models. However, there remain well-known difficulties in performing inference on complex, nonlinear systems. From a user perspective, inference methods for such systems are typically computationally expensive or technically difficult to use \cite{he2010}.  Additionally, limitations in data collection, combined with complex model structures, may mean parameter estimates are not uniquely determinable, or \emph{nonidentifiable} \cite{raue2009,frohlich2014}. There is also the subtle problem of \emph{model misspecification} \cite{kennedy2001,brynjarsdottir2014,king2015avoidable}, where there are significant discrepancies between the model and the true data generating process. Such discrepancies can cause bias in inference and inaccuracies in forecasting if not considered carefully.

Many recent advances in inference methods incorporate uncertainty from misspecification into the modelling process as an additional stochastic process, under both likelihood/frequentist frameworks \cite{king2015avoidable} and Bayesian frameworks \cite{chatzilena2019}. However, this compounds the computational expense of the inference problem, due to the increased complexity. The core of the expense is the additional marginalisation step over the stochastic process model that this introduces to the calculation of the likelihood function. Even with state-of-the-art methods, these stochastic methods can often take at least five times longer than deterministic counterparts \cite{chatzilena2019}.

An alternative to explicit error modelling is to only approximately solve or enforce a deterministic model. This approach allows for violations of the deterministic model without entirely sacrificing the model-based information required for forecasting. This can help prevent large prediction errors from strict enforcement of a misspecified model. Only approximately solving the equations also reduces the computational expense, which is useful in forecasting where results are time-sensitive and resources limited.

We follow this approach here and utilise an existing statistical method based on this idea, called \emph{generalised profiling} \cite{ramsay2007}. We introduce a variant of this approach under a new statistical interpretation and describe how this assists in addressing nuisance parameters, model misspecification and uncertainty quantification. Our statistical interpretation differs from that of introducing an explicit stochastic process model; instead, we enforce the model as a stochastic constraint following ideas introduced in the statistics and econometrics literature \cite{durbin1953-wt,theil1961pure}. This can be thought of as a form of mixed estimation, or as a \textit{prior likelihood} \citep{edwards1969-vc} term. This provides a previously lacking explicit motivation for profiling (maximising) out nuisance parameters in the style of profile likelihood instead of marginalising them out.

The rest of the article is laid out as follows. In \Cref{sec:bg}, we describe the general model fitting problem in more detail. In \Cref{sec:methods}, we describe our methods, including a variant of generalised profiling. In \Cref{sec:casestudy}, we apply the methods to a case study of a measles outbreak in Samoa, where the data is non-ideal and some classical methods break down. Finally in \Cref{sec:discussion} we discuss benefits and limitations to the approach.

\section{Background}\label{sec:bg}

Mechanistic models in biology are often deterministic dynamical systems, where the evolution of state of the system, denoted by $x$, is defined either in discrete time, $x_{t+1} = F(x_{t};\theta)$, or more commonly as a set of differential equations:
\begin{equation}
    \frac{dx}{dt} = f(x;\theta),
\end{equation}
where $f$ is some vector field, and $\theta$ its parameters. The state $x$ itself may also be discrete or continuous-valued.

The traditional method of parameter estimation is based on minimising some measure of the distance $d(y,x(t;\theta))$ between the data, $y$, and the parameter-dependent state of the model, $x(t;\theta)$, under some observation model $g$, for example:
\begin{equation}
    d(y,x(t;\theta)) = \lVert y - g(x(t,\theta)) \rVert^2_2.
\end{equation}
The above corresponds to a typical `least squares' formulation and has a natural (though not necessary) interpretation under the assumption of normally-distributed measurement errors. This formulation has a few difficulties, however \cite{ramsay2017}. Firstly, it may be computationally expensive to evaluate $x$ since this is usually carried out by numerical integration, especially in the presence of stiffness (time-scale separation) or other related phenomena. Compounding this, estimating parameters by minimising this distance using iterative methods means numerical integration must usually be carried out repeatedly for different parameter values. Furthermore, many models have regions of parameter space where they exhibit unexpected or computationally expensive behaviour.

This formulation also implicitly assumes that the models $f$ and $g$ are ``correct'', in the sense that the system that generates the data $y$ is exactly described by $f$ and $g$. This may not be true: the models may be \textit{misspecified}. This causes problems with biased estimates of the parameters and state \citep{brynjarsdottir2014}. One way to model this is to use an explicit stochastic process, for example a Gaussian process, to capture the misspecification \citep{kennedy2001}. This adds another stochastic term to the model, which is then used to account for discrepancies between the best-fitting model and the data. Similarly, stochasticity can be incorporated directly into the dynamics of the model. A typical continuous-time, continuous-state model is then a stochastic differential equation (SDE) of the form:
\begin{equation}
    dx = f(x;\theta) dt + \sigma(x;\theta)dW_t,
    \label{eq:sde}
\end{equation}
where $W_t$ represents a Wiener process \citep{Van_Kampen1992-ik,law2015data}. The error represented by the second term can be interpreted as either inherent stochasticity or as representing a form of misspecification due to missing influences, though these interpretations are subject to a number of subtleties \citep{Van_Kampen1992-ik}. 

Regardless of how the stochastic model is interpreted, inference on such models is significantly more expensive than the deterministic ODE system. The state $x$ is typically unobserved, and the probability distribution used in the likelihood function $p(y | \theta,\sigma)$ thus requires marginalising over this. Much of the recent literature on estimating the parameters of complex dynamic models hence aims at developing more efficient ways of marginalising over latent states to evaluate this form of the likelihood or some approximate replacement for it. These approaches usually either exploit the partially-observed Markov process/state-space formalism or use simulation-based inference \cite{diggle1984monte,breto2009time,wood2010statistical,Hartig2011-ot,fasiolo2016comparison,king2016statistical,Ionides2017-oa,breto2018modeling,chatzilena2019}. Discrete-time, and/or discrete-state, stochastic models are often used for modelling complex systems in biology and statistics \citep{wilkinson2018stochastic,durbin2012time}, and these introduce similar computational burdens.

\section{Methods}\label{sec:methods}
\subsection{Generalised Profiling}\label{sec:gp}

An alternative to explicitly representing the model misspecification is to weakly, or approximately, enforce the model. We do this by using the generalised profiling method introduced by \citet{ramsay2007}. The generalised profiling method, also known as the parameter cascade method, was built on the methods of functional data analysis \citep{ramsay2005-qw}, in order to incorporate functional assumptions about the underlying data in the form of differential equations. We show that, under a stochastic constraint interpretation, this can also be used to approximately enforce a differential equation model and hence allow for misspecification.

\subsubsection{Standard formulation}
In its classical formulation \citep{ramsay2007,ramsay2017}, the generalised profiling method is comprised of two nested optimisation problems, extending the two-stage nonlinear least squares algorithm. The inner problem is a smoothing problem, regularised by the model for some given fixed $\theta$.

\begin{argmini}|s|
    {c | \theta}
    {\lVert y-g(x) \rVert ^2_2 + \lambda \int \left(\frac{dx}{dt} - f(x; \theta)\right)^2 dt,}
    {\label{eq:inner-obj}}
    {\hat{c}(\theta) = }
    \addConstraint{x}{=\Phi c.}
\end{argmini}
Here the state $x$ is represented in terms of some basis comprising the columns of $\Phi$, which is typically a basis of B-splines. This inner problem is tuned with the hyperparameter $\lambda$, which acts as a tradeoff between interpolation (when $\lambda$ is low) and model matching (when $\lambda$ is high). This produces some optimal value for $c$ for the given $\theta$, which we denote $\hat{c}(\theta)$. This value is then passed into an outer optimisation problem
\begin{mini}|s|
    {\theta}
    {\lVert y- g(\Phi \hat{c}(\theta)) \rVert^2_2,}
    {\label{eq:outer_obj}}{}
\end{mini}
which performs the parameter optimisation. Because the inner objective \Cref{eq:inner-obj} uses a spline representation to perform state estimation, we can avoid integration by collocating at a chosen set of points in the time domain. The derivatives in the integral term can be directly determined by differentiating the spline representation. In contrast to standard derivative matching, this hybrid approach allows for unobserved states to be dealt with and simultaneously estimated.

The generalised profiling approach is closely related to spline smoothing and other forms of non-/semi-parametric regression. In particular, the inner problem can be seen as a generalisation of standard smoothness penalties to differential equation penalties building on the functional data analysis literature \citep{ramsay2005-qw,ramsay2017}. As discussed by \citet{wahba1990spline,ruppert2003}, the solution to such smoothing problems can often be interpreted as a form of Bayes estimate or as a predictive estimate for random-effects $c$. Similar points were raised in the discussion of \citep{ramsay2007}. \citet{ramsay2007,ramsay2017}, however, stress that they do not view their estimates as arising from an explicit stochastic process model. One aim of the present work is to further develop the interpretation of the work done by \citet{ramsay2007}.

Also of note, both conceptually and computationally, is that the `nuisance' parameters $c$ are maximised out to give $\hat{c}(\theta)$, rather than averaged out (which would be expected for random effects). This act of replacing the nuisance parameters with their maximum likelihood estimates, for each value of the interest parameter, is known as \textit{profiling} \cite{pawitan2001all}. Hence, it appears that the outer objective is intended as the profile likelihood of the inner objective, with respect to $\theta$, though this is not made explicit in the original paper by \citet{ramsay2007}. However, if we are to take this interpretation, questions arise as to why the outer objective is different to the inner objective.
We next consider a reformulation of the profiling approach which we feel more naturally unifies the inner and outer optimisation problems and, also leads to an explicit justification for profiling $c$ out. We first motivate this in the linear case.

\subsubsection{Reformulation: linear case}
Consider the linear regression problem for data $y$ of the form 
\begin{equation}
    y = G\Phi c + e,
    \label{eq:regress}
\end{equation}
where $G$ is a linear observation operator, $\Phi$ consists of a basis with a large number of basis functions (`columns' of $\Phi$) relative to the number of observations (`rows' of $G$), and $e$ is observational noise with mean $0$ and covariance $\Gamma$. As stated this problem typically requires regularisation of some form to constrain the $c$. Typical statistical interpretations of such regularisation are in terms of Bayesian models \cite{wahba1990spline,ruppert2003} or, in non-Bayesian inference, in terms of random-effects models \cite{ruppert2003}. In both cases the $c$ coefficients are taken to be random. 

An alternative way to introduce a similar regularising effect based on external or `prior' information with a statistical interpretation, but in which $c$ is non-random, was introduced by \citet{theil1961pure}, building on the work of \citet{durbin1953-wt}. This amounts to considering, in addition to \Cref{eq:regress}, the additional linear regression equation for another observable random variable $r$:

\begin{equation}
    r = H\Psi c + \nu.
    \label{eq:regress-prior}
\end{equation}
Here $H$ is a linear observation operator, $\Psi$ consists of a potentially different set of basis functions, and $\nu$ is observational noise with mean $0$ and covariance $\Sigma$. This will be assumed to be independent of $e$, representing an independent source of information on $c$. In this context $c$ is assumed fixed (non-random), while $y$ and $r$ are random. The \citet{theil1961pure} procedure amounts to assuming we have observations of both random variables, i.e. observations of $y$ and $r$, and carrying out a standard analysis given these observations. Here the observation of $r$ plays the role that the prior or overall mean typically plays in Bayesian or random effects models. \citet{theil1961pure} call this \textit{mixed estimation}, though this is to be distinguished from mixed modelling in which $c$ would typically be random. This is closely related to the concept of a \textit{prior likelihood} introduced by \citet{edwards1969-vc}, which is additional information in the form of a likelihood for a fixed parameter such as $c$ based on real or hypothetical prior data on a random variable such as $r$. A recent treatment of the method of stochastic restrictions in the sense of \citet{theil1961pure} can be found in \citet{rao2008linear}. H-likelihood \citep{lee1996hierarchical,lee2018generalized} is another related, though distinct, idea. 

The above equations can be `stacked' to give

\begin{equation}
\begin{bmatrix}
y \\
r 
\end{bmatrix}
= 
\begin{bmatrix}
G\Phi\\
H\Psi
\end{bmatrix}
c
+ 
\begin{bmatrix}
e\\
\nu
\end{bmatrix}
\label{eq:regress-stacked}
\end{equation}
This can be solved using the generalised least squares method, using the notation $|| f||_A^2 \coloneqq f^TAf$
\begin{equation}
        \min_{c}\ \ \lVert y-G\Phi c \rVert ^2_{\Gamma^{-1}} + \lVert r-H\Psi c \rVert ^2 _{\Sigma^{-1}},
        \label{eq:least-squares-stoch}
\end{equation}
which has the usual explicit solution
\begin{equation}
\hat{c} = (\tilde{G}^T\tilde{G})^{-1}\tilde{G}^T\tilde{y},
\end{equation}
where $\tilde{G} = \begin{bmatrix}
LG\Phi\\
MH\Psi
\end{bmatrix}$, $\tilde{y}=\begin{bmatrix}
Ly \\
Mr 
\end{bmatrix}$, where $L$ and $M$ are whitening matrices of $\Gamma$ and $\Sigma$ respectively ($L^TL = \Gamma^{-1}, M^TM=\Sigma^{-1}$). $\tilde{G}^T\tilde{G}$ is invertible if the intersection of the nullspaces of $LG\Phi$ and $MH\Psi$ is the empty set \citep{engl_regularization_1996}.

We now consider the case where the auxiliary information of \Cref{eq:regress-prior} takes the form of a linear differential equation. We first write this in operator form in the exactly specified case as
\begin{equation}
r = L(\theta)x
\label{eq:r-de}
\end{equation}
where $L(\theta) = \mathcal{D}-A(\theta)$ represents the autonomous system $\mathcal{D}x = f(x;\theta) = A(\theta)x$, where $A(\theta)$ is a linear operator acting on $x$ that depends on parameters $\theta$, and $r$ represents external forcing terms not depending on $x$. We have written $r$ on the left-hand side to mimic \Cref{eq:regress-prior}. If we instead only enforce this as a \textit{stochastic constraint}, for stochastic $r$, analogously to \Cref{eq:regress-prior}, we obtain the least squares problem 
\begin{mini}|s|
        {c|\theta}
        {\lVert y-G\Phi c \rVert ^2_{\Gamma^{-1}} + \int \left(m\left( r-L(\theta)\Phi c \right)\right)^2 dt}
        {\label{eq:linear-least-squares-stoch}}{}
\end{mini}
where, in comparison to the regression formulation \Cref{eq:least-squares-stoch}, we have taken $\Psi = L\Phi = (\mathcal{D}-A)\Phi$ and have enforced the stochastic constraint in the continuous limit with weighting function $m(t)$ (this is solved discretely in practice, however). The notation ${\displaystyle \min_{c|\theta}}$ indicates that the above is a least-squares problem for $c$ given each choice of $\theta$.

The infinite-dimensional nature of the differential equation constraint requires care. A stochastic interpretation of this infinite-dimensional constraint can be given by interpreting it as enforcing \Cref{eq:r-de} stochastically, i.e. as enforcing
\begin{equation}
r = L(\theta)x + \Sigma DW,
\label{eq:r-de-gen}
\end{equation}
where $DW$ stands for the (formal) time derivative of a multi-dimensional Wiener process \citep{law2015data}. To ensure a consistent interpretation, when this is implemented discretely as a set of regression equations in terms of `observations' of the model, we use an Euler-Maruyama discretisation scheme \citep{kloeden2013numerical} for the derivatives. This amounts to taking, for the $i^\mathrm{th}$ `observation' (discretisation grid point) of the equation,
\begin{equation}
\begin{aligned}
(Dx)_i &\approx \frac{x_{i+1} - x_i}{\Delta t}\\
(DW)_i &\approx \frac{1}{\sqrt{\Delta t}}e_i,\ \ e_i \sim \mathcal{N}(0,I).
\end{aligned}
\end{equation}
Multiplying this through by $\Delta t$ to put it in differential form gives the standard Euler-Maruyama discretisation. 

The objective \Cref{eq:linear-least-squares-stoch} reduces to the standard inner objective for generalised profiling of \Cref{eq:inner-obj} when the differential equation is linear and, in particular, when our \textit{stochastic prior information consists of the `observation' $r=0$}. We will assume this in general, which is analogous to taking the overall mean in the Bayes/random effects formulations as zero, but note that this is not a necessary assumption and could be modified to allow for (random observations of) non-zero forcing terms. Importantly, \textit{maximising} out $c$ for each $\theta$ corresponds to \textit{profiling} a standard joint likelihood function $\mathcal{L}(\theta,c;y,r)$ under appropriate conditions.

\subsubsection{Reformulation: general case and link to standard formulation}
We now reformulate the general nonlinear problem in terms of a single overall objective, $l(\theta, c)$:
\begin{mini}|s|
    {\theta, c}
    {l(\theta, c) \,&= \lVert y-g(x) \rVert ^2 _{\Gamma^{-1}} + \int \left(m\left(Dx - f(x; \theta) - r\right)\right)^2 dt}
    {\label{eq:general-gp}}{}
    \addConstraint{x}{=\Phi c}
\end{mini}
We will in general take $r = 0$ and hence drop this from now. As discussed above, we can approximate the second term using an Euler-Maruyama discretisation, for some choice of time step $\Delta t$, and write:

\begin{mini}|s|
    {\theta, c}
    {l(\theta, c) \,&= \lVert y-g(x) \rVert ^2 _{\Gamma^{-1}} + \lVert Dx - f(x; \theta)\rVert^2_{\widehat\Sigma^{-1}}}
    {\label{eq:raw_obj_fn}}{}
    \addConstraint{x}{=\Phi c}
\end{mini}
where $\widehat{\Sigma} = \Sigma\Delta t$. 

We can also add additional prior likelihood terms in the same manner, alongside the differential equation term. We consider additional covariance terms in more detail in the following subsection. First we relate the above version to the standard generalised profiling approach.

As indicated above, this formulation effectively replaces the objective of the outer optimisation problem in \Cref{eq:outer_obj} by a \textit{profile} of the overall objective over a vector `nuisance' parameter $c$, in the same sense as profiling (negative log-)likelihoods \citep{pawitan2001all}. For example, denoting the overall objective by $l(\theta,c;\Gamma, \Sigma)$, for fixed $\Gamma, \Sigma$ and for the constraint $x = \Phi c$ imposed, our problem can be decomposed into two analogous stages: 
\begin{equation}
    \min_{\theta}\ \min_{c | \theta}\ \ l(\theta,c;\Gamma, \Sigma).
\end{equation}
While the inner optimisation step is the same here as the standard formulation \Cref{eq:inner-obj}, our outer optimisation retains the model misfit term. By defining 
\begin{equation}
        l_p(\theta,\hat{c}(\theta));\Gamma, \Sigma) = \min_{c | \theta}\ \ l(\theta,c;\Gamma, \Sigma),
\end{equation}
our outer optimisation then corresponds to 
\begin{equation}
    \min_{\theta}\ \ \ l_p(\theta,\hat{c}(\theta));\Gamma, \Sigma).
\end{equation}

Existing literature using the generalised profiling method \cite{cao2007,ramsay2007,hooker2011,campbell2013,xun2013} appears to focus on parameter inference, and neglect the predictive properties of the method. Some work has been done in prediction in \cite{hooker2011}, but in the context of tuning the smoothing hyperparameter $\lambda$ of the classical formulation. This was done by integrating the model exactly from a set of points chosen on the approximated state. This integration was never extended past the time horizon of the data, meaning that true prediction was never done. Further, it is noted that depending on the choice of $\lambda$, the integral of the model and the estimated state may not necessarily agree. Thus this raises the question of how to utilise and choose information gained from the inference process for prediction. This was also a key question raised by the discussants of \citet{ramsay2007}.
To address this, we introduce a natural method for performing prediction using our objective function. By extending the fitting time-window beyond that supplied by the data, the overall objective naturally performs extrapolation of the data as informed by the model. This is due to the model-misfit term, which imposes the trajectory's behaviour further forward in time, while respecting the information from the data near the regions where it is available. Practically, this means that prediction can be done as part of the inference process, with no further decisions required. This is another benefit to viewing the inner and outer optimisation problems as simply different profiles of the same overall objective.

\subsubsection{Covariance Estimation and Iterative Solution Process}\label{sec:tuning}

In the classical formulation, the tuning of the hyperparameter $\lambda$ can be done efficiently in the linear case, through the application of the generalised cross-validation criterion \cite{wahba1990spline}, among other, typically prediction-oriented, data-driven methods. However, when the model $f$ becomes nonlinear, then the approaches are typically limited to grid search methods \cite{ramsay2017,campbell2013}.

In our reformulation, the estimation of the covariances of the errors are analogous to the tuning of $\lambda$.
Hence, unlike the standard least squares form of generalised profiling, here covariance-related terms in the negative log-likelihood are not dropped from the final form of our objective. Following above assumptions about the errors in the model, we get for Gaussian errors the following negative log-likelihood objective function.
\begin{equation}
    l(\theta, x) = \lVert y-g(x)\rVert^2_{\Gamma^{-1}} + \lVert Dx - f(x;\theta) \rVert^2_{\widehat\Sigma^{-1}} - \frac{1}{2}\log|\Gamma| - \frac{1}{2}\log |\widehat\Sigma|
    \label{eq:log-likelihood}
\end{equation}
However, the concurrent estimation of both the parameters and covariances is difficult due to computational issues with convergence, as well as robustness concerns. Such problems are typically resolved by using iterative methods, such as generalised least squares methods \cite{carroll1988}, which alternate between estimation of the parameters and covariances. These are early-stopping versions of iteratively-reweighted least squares, but the optimal number of iterations $N$ to take are hard to determine in most cases, and empirical experiments done by \citet{carroll1988} show that $N \geq 2$ is advised. We also observe that in our formulation, there is a potential degeneracy in the model fit term, due to it being able to be exactly satisfied, leading the estimated model fit covariance to unboundedly decrease. We counter this behaviour by setting restrictions on the maximum or minimum estimated covariances of each term.
We also note that this iterative procedure can be used in the form of quasi- or pseudo-likelihood estimation where, for example, the estimates of the mean (state) function parameters are based on quasi-likelihood/generalised least squares procedures, while estimates of the covariances use (typically normal-theory) maximum likelihood or related methods \cite{carroll1988,ruppert2003}. This allows the use of non-normal likelihood expressions, at least for estimating the parameters of the mean (state) function. The iterative process used is given in \Cref{alg:irls-gp}, and we expand on the details of the implementation in the Supplementary Material.

\begin{algorithm}[h]
    \caption{Iteratively reweighted least squares algorithm}
    \label{alg:irls-gp}
    \DontPrintSemicolon
    \SetKw{Break}{break}
    \KwData{initial iterate $x_0$, initial weights $w_0$, negative log-likelihood function $l(x, w)$, maximum iterations $N$, error threshold $\epsilon$, minimum weight $w_{min}$, maximum weight $w_{max}$}
    \KwResult{optimal values $\hat x$, optimal weights $\hat w$}
    \Begin{
    initialisation\;
    $f_0 \gets l(x_0, w_0)$\;
    \For{$i=1$ \KwTo $N$}{
      $x_i \gets \arg\min_x l(x, w_{i-1})$\;
      $w_{prop} \gets \arg\min_w l(x_i, w)$\;
      $w_i \gets \max(w_{min}, \min(w_{prop}, w_{max}))$\;
      $\Delta f_i \gets f_i - f_{i-1}$ \;
      \If{$\Delta f_i < -\epsilon$}{\nllabel{algln:begincheck}
          \tcp{divergence}
          $\hat x \gets x_{i-1}$\;
          $\hat w \gets w_{i-1}$\;
          \KwRet{}\nllabel{algln:endcheck}
      }
    }
    $\hat x \gets x_i$\;
    $\hat w \gets w_i$\;
    }
\end{algorithm}

\subsection{Uncertainty Quantification}\label{sec:uq}
Next, we consider statistical uncertainty quantification, in the sense of forming appropriate uncertainty intervals with (approximate) coverage properties.

In the classical approach, the intervals can be extracted by treating the coefficients $c : x=\Phi c$ as nuisance parameters that depend on the model parameters $\theta$. The outer objective function then allows the uncertainty to be propagated from the data space $y$ to parameter space, via a linearisation. This leads to Wald-style confidence intervals; however, if the underlying log-likelihood is not well approximated by a quadratic, then this approach can give misleading results \cite{pawitan2001all}. 

An alternative way to perform analysis on the uncertainty of quantities of interest is using the profile likelihood \citep{pawitan2001all,Ionides2017-oa}. We can apply typical frequentist sampling theory methods to form approximate likelihood-based confidence intervals for particular quantities of interest from their profile likelihoods. For a scalar quantity of interest, this is typically done by choosing a cutoff point $k$ such that 
\begin{equation*}
    k = \exp\left(-\frac{1}{2} \chi^2_{1,(1-a)}\right)
\end{equation*}
where $\chi^2_{1, (1-a)}$ is the $(1-a)$-th percentile of the chi-squared distribution with 1 degree of freedom. This cutoff then defines an asymptotic $100(1-a)\%$ confidence interval (under appropriate regularity conditions \cite{pawitan2001all}):
\begin{equation}
    \left\{\omega, \frac{\mathcal{L}_p(\omega)}{\mathcal{L}_p(\hat\omega)} > k \right\}
\end{equation}
where $\omega$ is the quantity of interest, $\mathcal{L}_p(\omega)$ is the profile likelihood of $\omega$, and $\mathcal{L}_p(\widehat\omega)$ is the profile likelihood of $\omega$ evaluated at the MLE, for which $\omega=\widehat\omega$. This confidence interval can also inform the identifiability of the quantity of interest \citep{raue2009,simpson2020}. This process can be carried out by re-solving the maximum likelihood estimation procedure over a grid of values for $\omega$, with an additional equality constraint for those values of $\omega$. We stress here that $\omega$ is not necessarily a model parameter, but can be any function of the model parameters $\theta$ or state variables $c$.

Another approach is to use a bootstrap-style sampling distribution \citep{davison1997bootstrap,efron1979bootstrap}, determined by repeatedly solving the (discretised) optimisation problem:
\begin{mini}|s|
    {\theta, c}
    {\lVert y-g(x) + e \rVert ^2 _{\Gamma^{-1}}&+ \lVert Dx - f(x; \theta) + \nu \rVert^2_{\widehat\Sigma^{-1}}}
    {\label{eq:rto-eqs}}{}
    \addConstraint{x}{=\Phi c}
\end{mini}
for a series of realisations of $e$ and $\nu$, which have zero mean and covariances $\Gamma$ and $\widehat\Sigma$. This can be done efficiently using automatic differentiation, and can be carried out in perfect parallel. 
In the inverse problems literature, the above approach falls under the umbrella of Randomised Maximum Likelihood (RML) \citep{oliver1996-ro}
methods that include randomised Maximum a Posteriori (rMAP) \cite{wang2018rmap} and Randomize-then-Optimize (RTO) \cite{bardsley2014}.
These methods are usually considered approaches to determining approximate Bayesian posteriors, or proposal distributions for sampling Bayesian posteriors.
In contrast, we view our approach as a way of determining a sampling distribution for the maximum likelihood estimator under prior information in the form of stochastic constraints in the style of \citet{theil1961pure}, or a prior likelihood in the style of \citet{edwards1969-vc}. Our interpretation of RML here is thus as a classical bootstrap method, and hence we will also refer to this as `RML bootstrap'.
Using the bootstrap interpretation, we can construct $100(1-2a)\%$ percentile interval \citep{efron1979bootstrap}:
\begin{equation}
    \left[ t^*_a, t^*_{(1-a)} \right]
\end{equation}
where $t^*$ are the bootstrap samples and $t^*_p$ is the $p$-th percentile of the bootstrap samples.

\section{Case Study}\label{sec:casestudy}

\begin{figure*}[t]
    \centering
    \begin{subfigure}[t]{0.475\linewidth}
        \includegraphics[width=\textwidth]{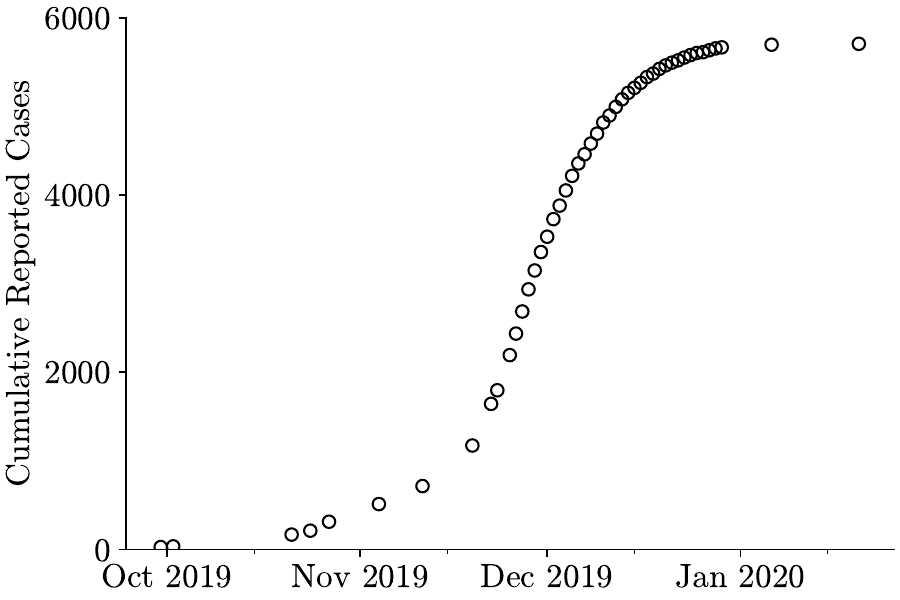}
    \end{subfigure}\hfill%
    \begin{subfigure}[t]{0.475\linewidth}
        \includegraphics[width=\textwidth]{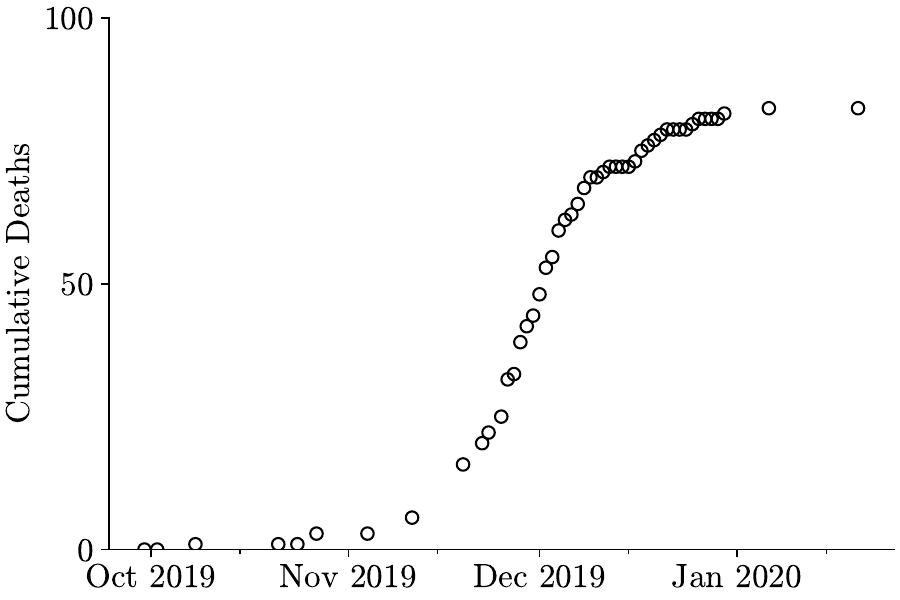}
    \end{subfigure}
    
    \medskip
    \begin{subfigure}[b]{0.475\linewidth}
        \includegraphics[width=\textwidth]{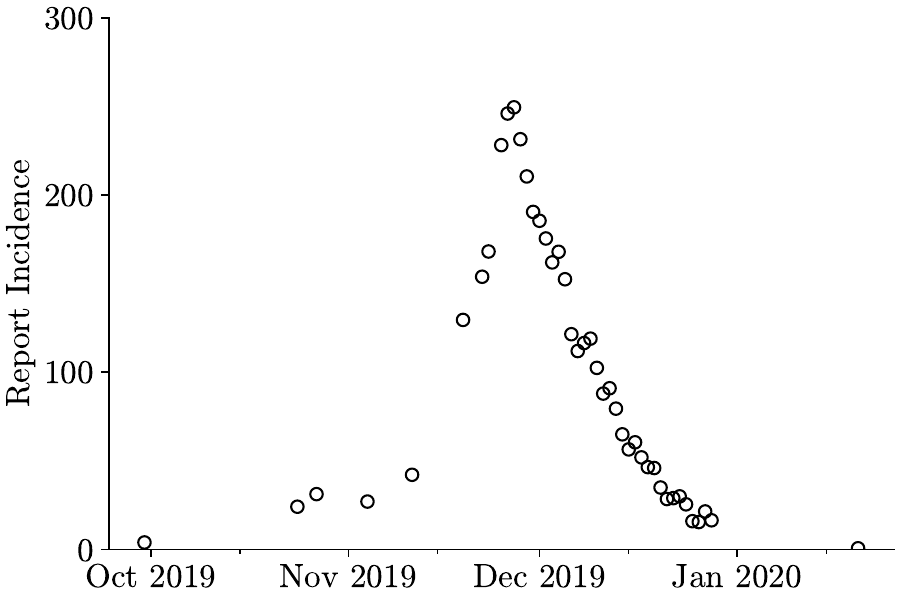}
    \end{subfigure}\hfill%
    \begin{subfigure}[b]{0.475\linewidth}
        \includegraphics[width=\textwidth]{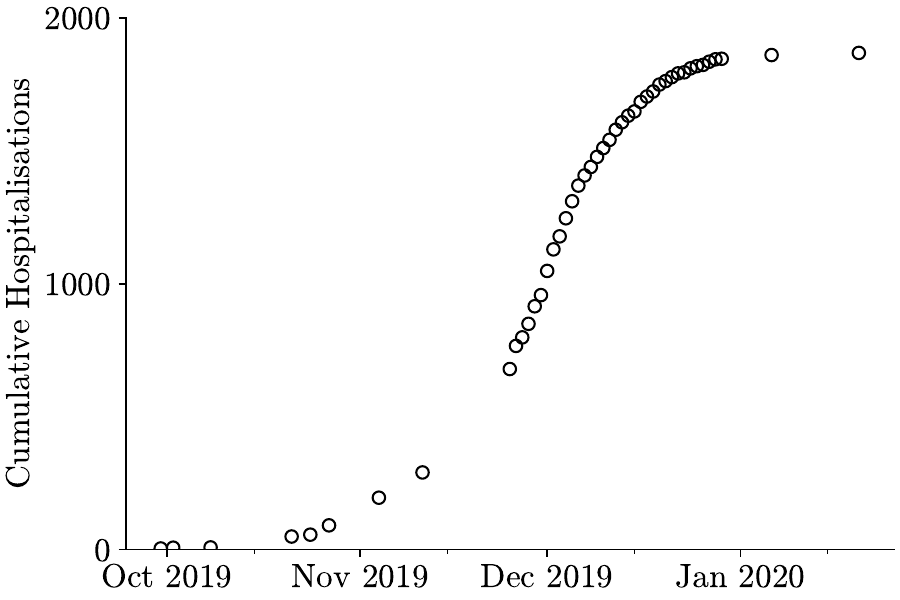}
    \end{subfigure} 
    \caption{Plots of the available data for the 2019 Samoan measles outbreak (current hospitalisations and cumulative discharges available, but not shown). Top-left: cumulative reported cases; top-right: cumulative reported deaths; bottom-left: report incidence (daily rate of new cases since last report); bottom-right: cumulative reported hospitalisations.}
    \label{fig:samoa_data}
\end{figure*}

We now consider an application of these ideas to a case study.
Between late September 2019 and early January 2020, the South Pacific nation of Samoa (population $\sim$200,000) experienced an unprecedented outbreak of measles. Data on the progression of the epidemic was released by various departments of the Government of Samoa in the form of press releases \cite{samoa2019}. Initially these were published by the Ministry of Health, but during the emergency period the function was performed by the National Emergency Operations Centre (NEOC). Initial press releases by the Ministry of Health were generally sparse, whereas NEOC press releases were daily. The data is presented in \Cref{fig:samoa_data}. Note that there is missing data for certain dates near the beginning of the outbreak.

In a preliminary effort, we used a less developed version of this method to make predictions for the Samoan measles epidemic, when contacted by the Samoan Observer in late November 2019 \citep{helen2019sciblogs, herald2019measles}. For our analysis here, we use data up to the 3rd of December to reproduce similar conditions. This is the point in time where two pertinent questions are raised: how long will the epidemic persist, and how bad will it be?

\subsection{Model}

To model the dynamics, we build upon the SEIR compartmental model, adding compartments for the additional data available (deaths and hospitalisations).

\begin{subequations}
\begin{align}
    \dot{S} &= -\beta S\frac{I}{N},\\
    \dot{E} &= \beta S\frac{I}{N} - \gamma E,\\
    \dot{I} &= \gamma E - (\alpha + \eta) I,\\
    \dot{R} &= \alpha I,\\
    \dot{H} &= \eta I - (\delta + \mu) H,\\
    \dot{G} &= \delta H,\\
    \dot{D} &= \mu H,\\
    \dot{I_c} &= \gamma E,\\
    \dot{H_c} &= \eta I.
\end{align}
\end{subequations}

$S$ represents the susceptible population, who become exposed ($E$) but not infectious at rate $\beta \frac{I}{N}$ on contact with the infectious ($I$). Exposed individuals become infectious $I$ at a rate $\gamma$, and report their infection (tracked by $I_c$). Infectious individuals can then recover to $R$ at rate $\alpha$, become hospitalised ($H$) at rate $\eta$. Hospitalisation is tracked by $H_c$, and hospitalised individuals recover and are discharged ($G$) at rate $\delta$, or die ($D$) at rate $\mu$. We also track the total number of reported cases ($I_c$), the total number of hospitalisations ($H_c$), and the total at-risk population $N = S+E+I+R+H+G$. We neglect vital dynamics, due to the relatively short duration of the outbreak, and also vaccination effects, which were implemented too late into the epidemic to have a significant effect. Pre-existing vaccinated individuals are modelled as a non-zero $R(t=0)$. Because we are assuming this existing population of vaccinated individuals, we define a critical threshold parameter $\Rz$ following \citep{vandenDriessche2008,vandenDriessche_Watmough_2002} as
\begin{equation}
    \Rz = \frac{\beta}{\alpha+\eta} \frac{S(t=0)}{N(t=0)}
\end{equation}

We note that not all states have corresponding data, that is we have a partially observed model. The observed states are described in \Cref{table:model_to_data}. Partially observed models introduce additional problems when fitting a differential equation model. At a basic level, it becomes difficult to quantify model misfit without estimating the state simultaneously. It also can introduce identifiability problems, depending on the state variables that can be observed \cite{tuncer2018}. In this analysis, we also relax the implicit assumption in \citep{tuncer2018,roosa2019} that the state at $t=0$ is known, in particular the ratio $S(t=0)/N(t=0)$.

\begin{table}
    \centering
    \caption{Data to state variable correspondences for all observable states of the measles model}
    \label{table:model_to_data}
    \begin{tabular}{c c}
        \hline
        Model State Variable & Corresponding Available Data \\
        \hline
        $I_c$ & Cumulative Reports\\
        $D$ & Cumulative Deaths\\
        $H$ & Current Admissions\\
        $H_c$ & Cumulative Hospitalisations\\
        $G$ & Cumulative Discharges\\
        \hline
    \end{tabular}
\end{table}

\subsection{Results}\label{sec:results}

\begin{figure*}[t]
    \centering
    \begin{subfigure}[t]{0.495\textwidth}
        \includegraphics[width=\textwidth]{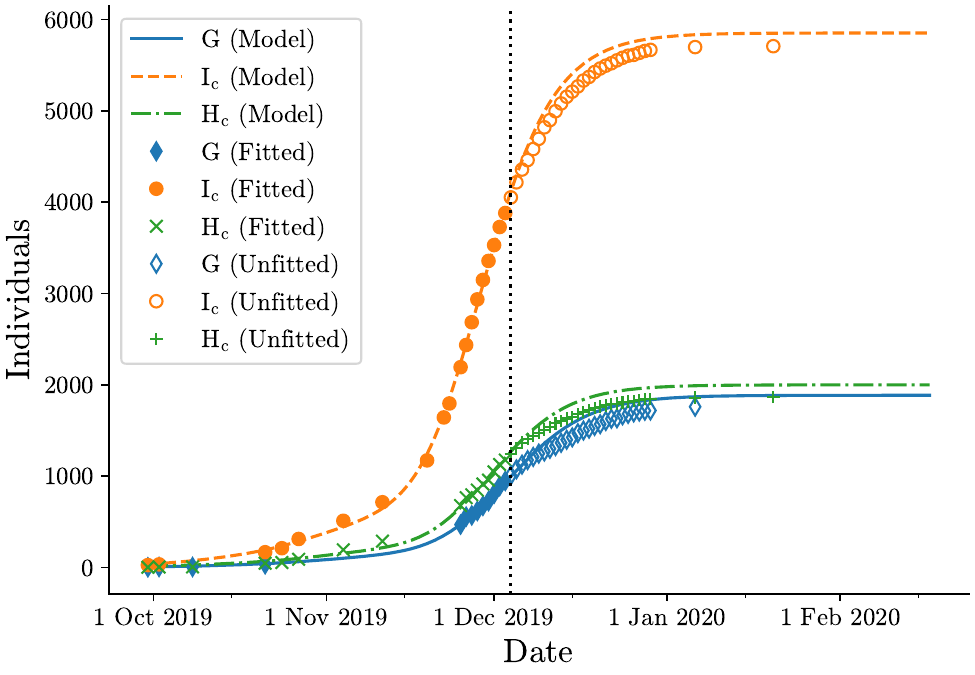}
        \caption{Cumulative reports ($I_c$), cumulative hospitalisations ($H_c$) and discharges ($G$)}
    \end{subfigure}\hfill%
    \begin{subfigure}[t]{0.495\textwidth}
        \includegraphics[width=\textwidth]{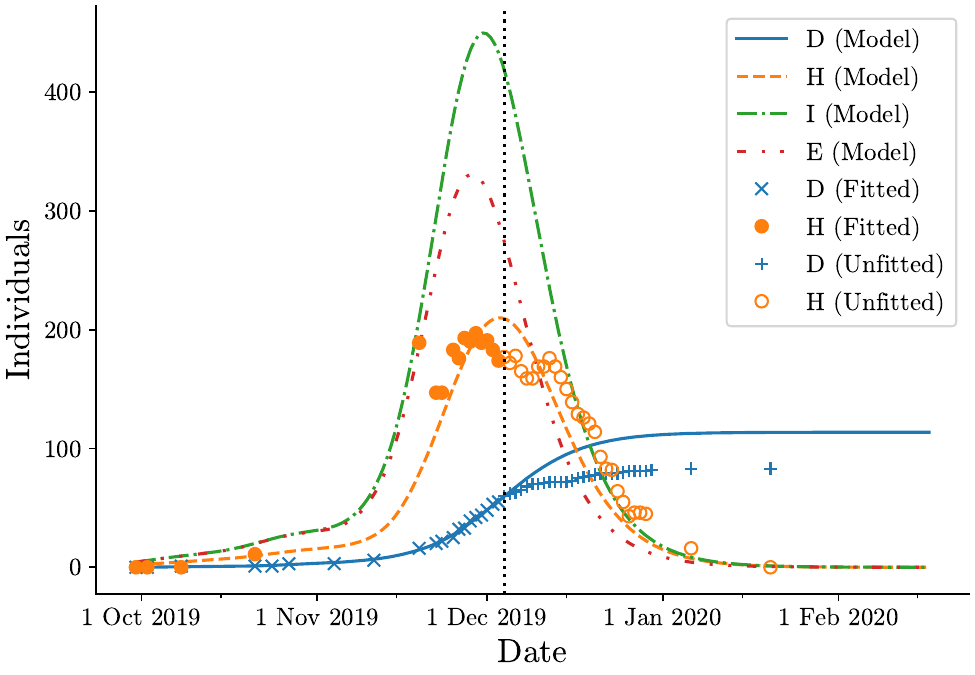}
        \caption{Deaths ($D$), incidence of non-infectious latent individuals($E$) and infectious individuals ($I$), and hospitalisations ($H$)}
    \end{subfigure}
    \caption{Recovered state estimates at $N=3$ iterations of the IRLS algorithm, which is the chosen early stopping point.}
    \label{fig:recovered_traj}
\end{figure*}

The methods were implemented in \texttt{Python} as described in \Cref{sec:gp}, and are initially validated on an SIR model with synthetic data (available in the Supplementary Material). The estimated trajectory of the Samoan measles outbreak, using data up to 3 December 2019, is presented in \Cref{fig:recovered_traj}.

We add two prior likelihood/regularisation terms that penalise the magnitudes of the parameters $\beta$ and $\gamma$. We tune the hyperparameters on the two penalty terms by analysis of synthetic data generated from the model (see the Supplementary Material for tuning and regularisation details). These terms penalise pathological behaviour where the $E$ or $I$ state estimates converge to small values where their derivative are small, as encouraged by the model misfit term.

We compute confidence intervals for the total reported cases and total deaths, as well as $\Rz$, using profile likelihood and RML bootstrap methods described above. The total reported cases and deaths are taken at the final time point of the state estimate, which is set to $1.25\times$ the final time point (when the epidemic was declared over by the Samoan Government), where dates are zeroed with respect to the date of discovery of suspected first case (30 September 2019). We draw 200 samples for the RML bootstrap method, and reject any samples that result in non-convergent optimisation behaviour ($n=22$). Intervals are computed for a 95\% confidence level, and a summary of results is presented in \Cref{tab:parameters_samoa}.

\begin{table}[htpb]
    \centering
    \caption{Point and interval estimates of quantities of interest for the Samoan measles case study.}
    \label{tab:parameters_samoa}
    \resizebox{\textwidth}{!}{%
    \begin{tabular}{l l c c c c c}
        \hline
        \multirow{2}{*}{\tabbox{Quantity}{Type}} & \multicolumn{2}{c}{Quantity} & \multirow{2}{*}{MLE} & \multirow{2}{*}{\tabbox{Profile}{Interval}} & \multirow{2}{*}{\tabbox{Bootstrap}{Interval}} & Truth \\
        & Interpretation & Symbol & & & & \\
        \hline
        Parameters & Infectivity & $\beta$ & 0.870 & -- & [0.813 -- 0.920] & --\\
        & Incubation & $\gamma$ & 0.717 & -- & [0.657 -- 0.772] &--\\
        & Recovery & $\alpha$ & 0.350 & -- & [0.316 -- 0.380] &--\\
        & Hospitalisation & $\eta$ & 0.176 & -- & [0.162 -- 0.186] &--\\
        & Discharge & $\delta$ & 0.306 & -- & [0.295 -- 0.318] &--\\
        & Death & $\mu$ & 0.0186 & -- & [0.0173 -- 0.0200] &--\\
        State Values & Total Cases & & 5785 & [5593 -- 6006] & [5621 -- 6043] & 5707\\
        & Total Deaths & & 111 & [102 -- 120] & [104 -- 120] & 83\\
        \tabbox{Other}{Quantities} & \tabbox{Reproduction}{Number} & $\Rz$ & 1.648 & [1.586 -- 1.717] & [1.596 -- 1.712] & -- \\ 
        \hline
    \end{tabular}%
    }
\end{table}

\begin{figure}[h]
    \centering
    \includegraphics[width=0.9\textwidth]{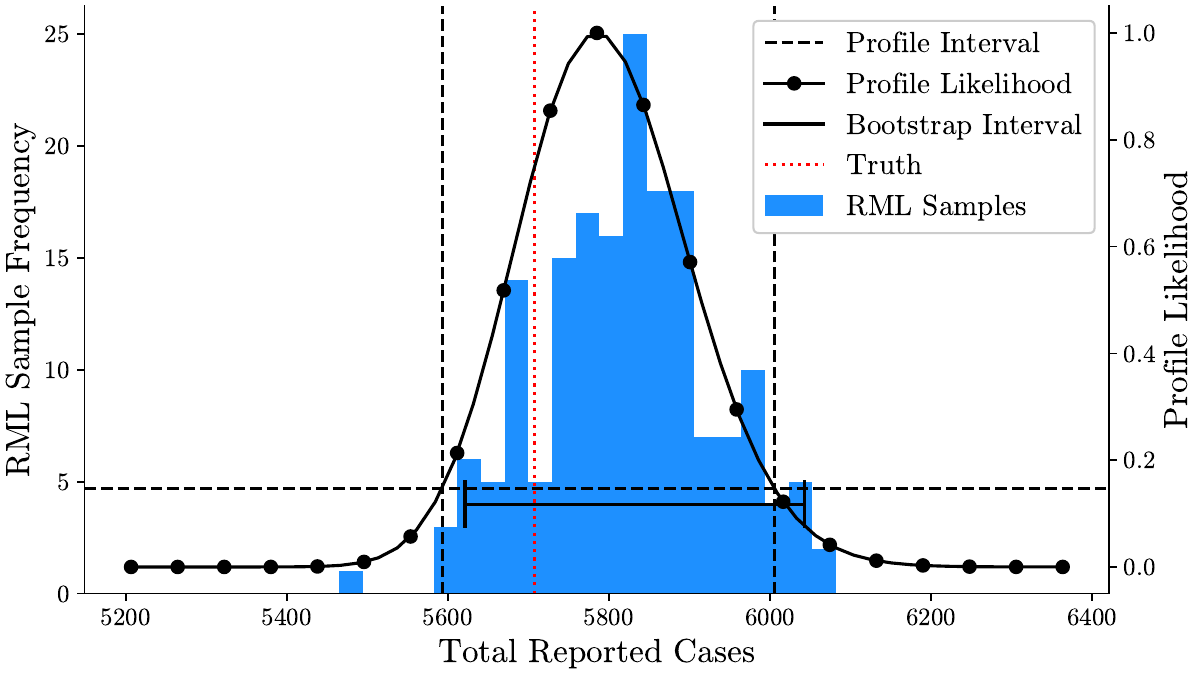}
    \caption{Profile likelihood and RML bootstrap samples ($n=178$) of the total reported cases, with corresponding 95\% confidence intervals marked.}
    \label{fig:final_cases_rml}
\end{figure}

\begin{figure}[h]
    \centering
    \includegraphics[width=\textwidth]{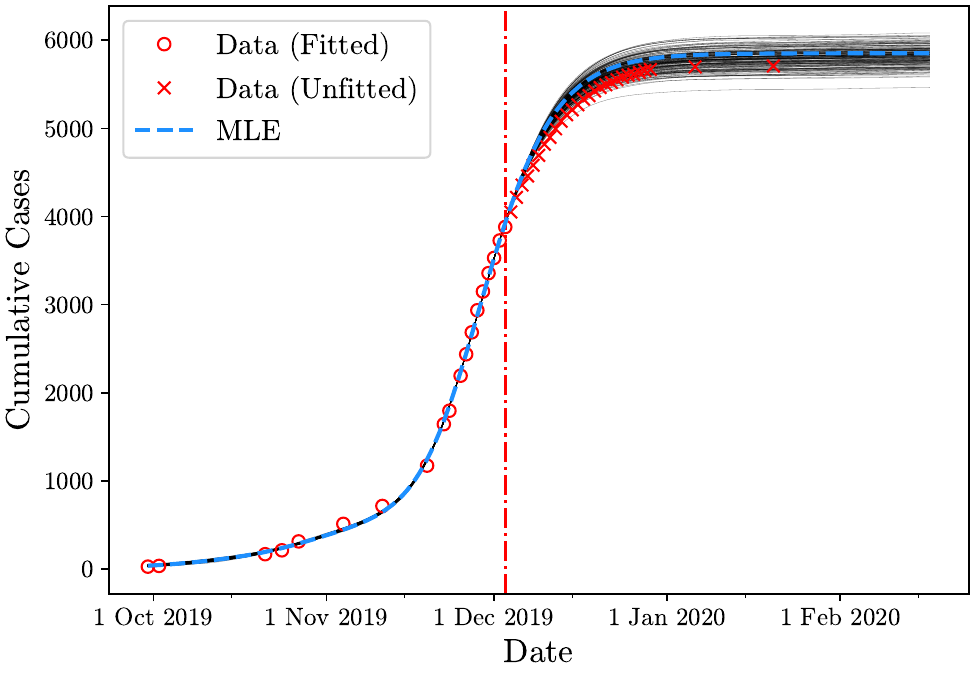}
    \caption{Timeseries of RML bootstrap samples of total reported cases ($n=178$).}
    \label{fig:total_cases_rml_est}
\end{figure}

We see that we capture the truth for the total number of cases with both profile likelihood and bootstrap intervals. The bootstrap sample distribution and profile likelilhood are presented in \Cref{fig:final_cases_rml}, and \Cref{fig:total_cases_rml_est} plots the RML state estimates for the total reported cases over time.

\begin{figure}[h]
    \centering
    \includegraphics[width=\textwidth]{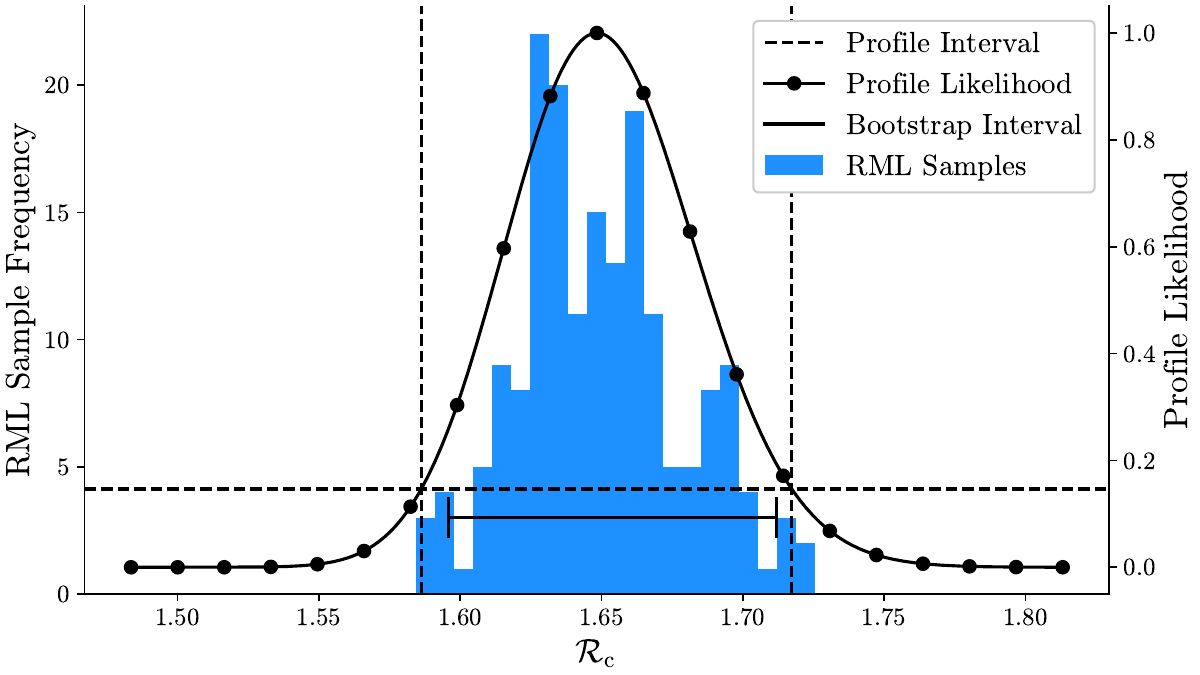}
    \caption{Profile likelihood and RML bootstrap samples ($n=178$) of $\Rz$, with associated 95\% confidence intervals marked.}
    \label{fig:r0profile}
\end{figure}

We see that the intervals for $\Rz$ by both methods roughly agree, though the bootstrap interval is slightly narrower, as shown in \Cref{fig:r0profile}.

\begin{figure}[h]
    \centering
    \includegraphics[width=\textwidth]{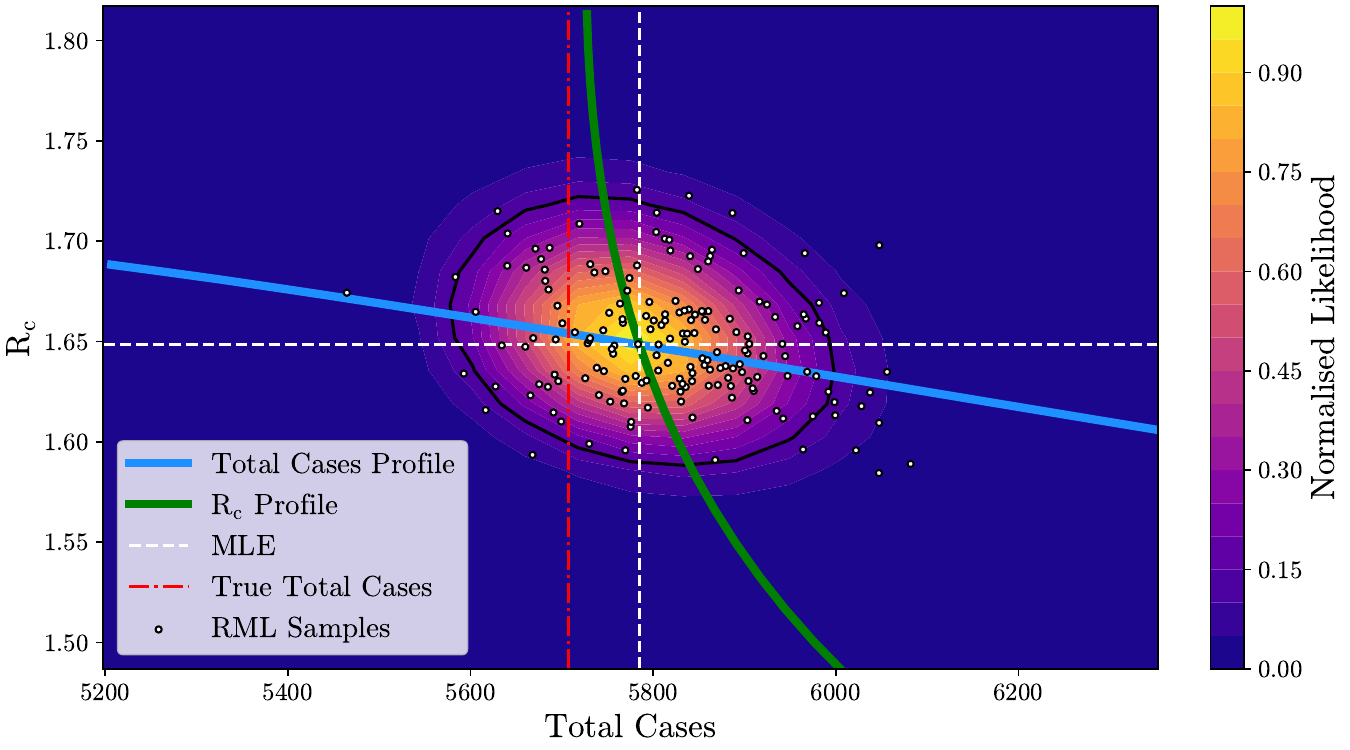}
    \caption{Joint profile of total cases and $\Rz$, with independent profiles of each variable and the RML samples overlaid. The contour highlighted in black represents the boundary of the 95\% confidence region, where the normalised likelihood value is approximately 0.15.}
    \label{fig:bivariate_profile}
\end{figure}

We also construct a joint profile of total cases and $\Rz$ in \Cref{fig:bivariate_profile}. We see that the RML samples and the 95\% confidence region agree. The univariate profiles of $\Rz$ and total cases follow the likelihood surface of the bivariate profile, as expected \citep{bolker_ecological_2008}.

\begin{figure}[h]
    \centering
    \includegraphics[width=\textwidth]{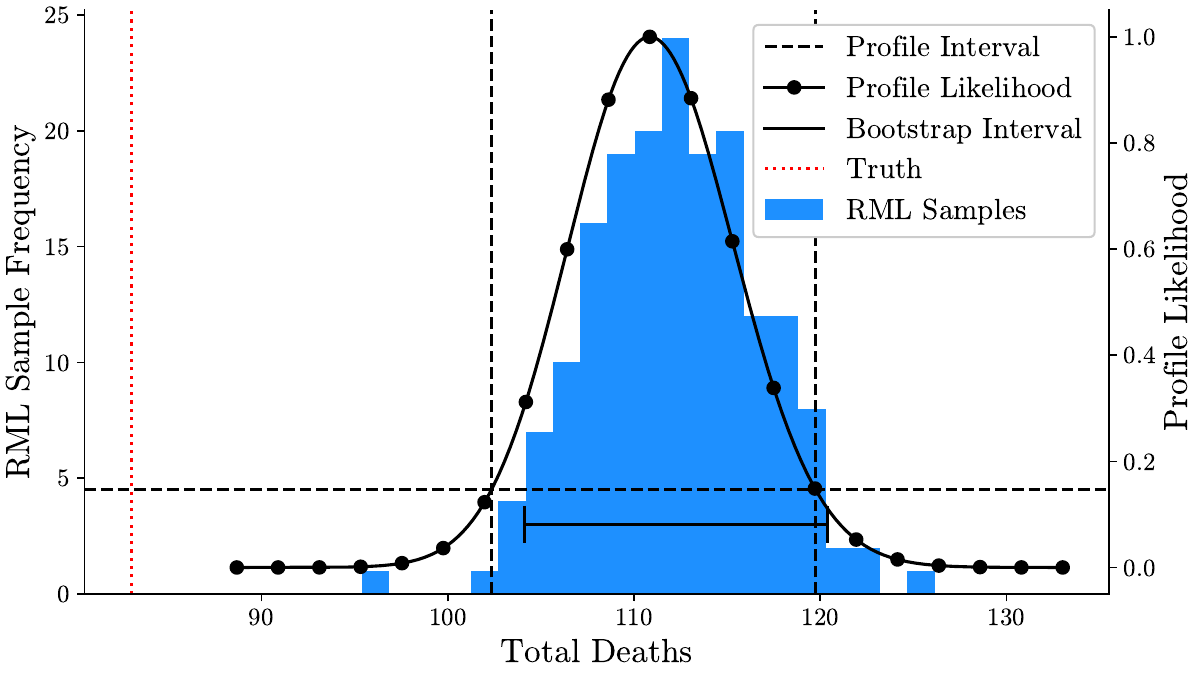}
    \caption{Profile likelihood and RML bootstrap samples ($n=178$) of the total deaths with associated 95\% confidence intervals marked.}
    \label{fig:mortality_bootstraps}
\end{figure}

We can also analyse the bootstrap samples for estimates of the death counts. We plot a histogram of the estimated deaths from the outbreak in \Cref{fig:mortality_bootstraps}, and derive an 95\% bootstrap confidence interval of [104 -- 120] and a profile interval of [102 -- 120]. We note that these bounds do not capture the true value of 83 deaths. This is likely due to the drop-off in deaths just outside of range of the provided data (see \Cref{fig:samoa_data}), which cannot be captured by a single mortality rate parameter.

\section{Discussion and conclusion}\label{sec:discussion}

The estimation and prediction of epidemics is a difficult task, in both theory and practice. In this paper, we have presented a likelihood-based reformulation of the standard generalised profiling method introduced and developed by \citet{ramsay2007}. This is based on ideas of mixed estimation in the sense of \citet{theil1961pure} and prior likelihood in the sense of \citet{edwards1969-vc}. This provides a natural rationale for profiling as a method of eliminating nuisance parameters and targeting interest parameters. We then use this to construct a framework for parameter and state inference of a model given data while allowing for model misspecification.

The standard formulation can be seen as a form of profiling, where the outer objective is the $\theta$-profile of the inner objective, albeit evaluated at $\lambda \to \infty$, where the model is exactly enforced. Our reformulation can thus be seen as more natural in a profile likelihood sense, as it interprets both stages as profiling a single log-likelihood containing both observational and model error. It also leads to the use of the profile likelihood \citep{pawitan2001all} and RML bootstrapping that allow us to perform uncertainty quantification on the parameters and states.

We have then applied these methods to a case study of a measles outbreak in Samoa, and shown that the estimates capture the truth for important quantities of interest, like $\Rz$ and the total number of cases. 

\subsection{Profile Likelihood Interpretation of Generalised Profiling}

A key methodological difference in our approach is in the treatment of the smoothing hyperparameter in the standard formulation in terms of covariance matrices of errors. This is similar to the ideas developed in the mixed-effects/random-effects/multi-level regression literature \citep{ruppert2003, hodges2013richly}, which embed spline models in mixed effect models using random effects. These approaches also relate closely to Bayesian interpretations and the work of \citet{wahba1990spline} on spline models. As noted above, however, a key motivation for our interpretation in terms of stochastic \textit{constraints}, rather than an explicit stochastic process model, is to provide a more natural statistical rationale for using profiling (maximisation) rather than marginalisation for eliminating nuisance parameters. Even in the fixed parameter setting, there is ongoing debate whether profiling or marginalisation better captures properties of likelihood when eliminating nuisance parameters \cite{berger1999, Aitkin_Stasinopoulos_1989, Aitkin_2010}.

When estimating covariances with our approach, there is still the open question of an optimal stopping criterion, or optimal number of cycles, to take in the IRLS algorithm. We find that even if the objective function terms are balanced, the solution tends to diverge without regularisation. In this case study, we opted to apply bound constraints on the weighting as a regularisation, but again the determination of the bounds is not clear. Furthermore, it is unclear how well covariance matrices can be estimated in this setting, reflecting similar difficulties noted in the mixed effects literature \cite{hodges2013richly}. Validation methods have been used with generalised profiling, such as the forwards prediction error suggested by \cite{hooker2010}, but there are problems when using this for ongoing or emerging outbreaks such as the Samoan measles data, since prediction is typically most sensitive to the more recent data. More recently, Bayesian approaches to mixed effects models often include additional information on covariances in the form of hyperpriors. \citet{huang2020} have considered this for their Bayesian reinterpretation of the standard formulation of generalised profiling, setting priors on the tuning parameter $\lambda$. In principle we could take an analogous approach in terms of our mixed estimation (stochastic constraints) formulation, but choosing appropriate forms for these additional terms faces similar challenges to those raised in specifying hyperpriors in mixed effect models \cite{hodges2013richly}. In general, estimating these hyperparameters remains a key challenge for all approaches. In this particular study, we choose bounds that lead to stable iterates in the maximum likelihood estimation procedure, and use a synthetic data study (see Supplementary Materials) to determine an appropriate magnitude for the parameter regularisation strength.

\subsection{Identifiability and Uncertainty Quantification}

Due to the ability to use the standard likelihood toolbox with the reformulation, we are also able to perform identifiability analysis in a straightforward manner. 

We see that $\Rz$ is identifiable in our analysis, which is consistent with the findings of \citet{roosa2019} that show that $\mathcal{R}_0$ is identifiable even if there are identifiability issues with the other parameters. We do note however, that the identifiability of the individual parameters is only possible due to the regularisation terms added, as prior likelihoods, to the likelihood function, and without them, there is evidence that there is practical nonidentifiability of model parameters $\beta$ and $\gamma$. This is likely related to the relaxation of the assumption that the initial state is known, since both are related to $\Rz$, where the initial condition appears (through $\dot{S}$ and $\dot{E}$).

Further, the coverage properties of the intervals we have constructed with regularisation are not well-known. Work in post-selection inference is formalising the properties of estimators that are regularised, or data-informed in some ways, in particular for methods like lasso \citep{hastie_statistical_2015}. There have also been alternative formulations that rely on imposing physically-motivated constraints on the problem to construct confidence intervals with appropriate coverage properties \citep{kuusela_2017}. In our work, we interpret the regularisation terms as prior likelihoods, or additional data.
This means the properties of our confidence intervals can be more directly justified in terms of the relevant joint sampling distribution involving both the present study data, and our prior regularisation data. This introduces a dependence on the quality of the prior data, similar to the dependence of Bayesian credible intervals on the prior distribution. %

Thus, there still remains the question of whether the approximate intervals we compute have close to the correct coverage, and how the sampling distribution should depend on the auxiliary data. The simple SIR experiment (in the Supplementary Material) suggests that the $\Rz$ profile should be skewed without the presence of prior regularisation data, whereas we see a more symmetrical profile when that additional regularisation data is added in the case study.

We find that that the estimation procedure is sensitive to initial guesses of the weightings (covariance matrices), especially for this data observation window, which ends around the peak of the epidemic. This manifests itself numerically as non-convergence after a large number of optimisation iterations.
In the Supplementary Material, we perform a synthetic study of observation windows for a standard SEIR model, and see that there is a sudden regime change in the estimation errors (for both state and parameters) once the observation window passes the turning point of the outbreak curve. 
This is consistent with the findings of \citet{wilke_bergstrom_2020} and \citet{roda2020}, which show that the predictive uncertainty window shrinks rapidly as data is collected past the turning point of the epidemic curve.
This is probably linked to the inability to determine the rate of recovery in the early stages of the epidemic, since the effect of recovery is small on the overall behaviour.

\subsection{Model Misspecification}

We see elements of model misspecification, which may stem from the ill-posedness of the problem. Empirically, we see that the fit of the total number of deaths, is biased in both the bootstrap and profile likelihood analysis. Drawing on our knowledge of the case study, we also know that our model does not capture the spatial characteristics of the spread in Samoa, or the fact that various non-pharmaceutical interventions were enacted in the form of curfews and a mass vaccination campaign. We could have allowed for this in the model, for example, by allowing for some of the parameters to be time-varying in a structured way, akin to how \citet{hooker2011} allow for a cyclical $\beta$ value in their endemic model. Further, we have information from previous measles outbreaks about the clinical presentation and infection characteristics of the virus, which could inform parameters such as $\gamma$ (the latent period) and $\alpha$ (the recovery rate). Although we have regularised $\beta$ and $\gamma$ towards 0 to make the estimation procedure better-behaved, this additional information used instead. In this way, this method can be seen as analogous to the Bayesian formulation, but instead of marginalising over the prior data, we maximise (profile) over the prior.

\subsection{Practical Considerations}

In our implementation of the method, we have used the \texttt{CasADi} framework \citep{andersson2019} and its interface in Python. We found that the estimation of the MLE for the Samoan measles case study took around 3 minutes. The profiles each take around 10 minutes to generate, dependent on the number of points to evaluate, and the 200 RML bootstrap samples take around 30 minutes. Most of these timings scale with the number of basis functions and collocation points, though the nonconvergence of the optimisation problem in many cases is more likely attributed to poor starting iterates. Our analysis of an SIR model (in the Supplementary Material) takes significantly less time, for a similar amount of data. Additionally, we have made a simplifying choice to perform all of the procedures in serial. However, there is room for parallelisation to significantly speed up the process of bootstrapping in particular, since bootstrap samples are independent of each another. For profiling, many different values of interest can be computed in parallel to each other, but the profile itself is constructed successively away from the MLE. This is due to there being limited regions of convergence for any given optimisation problem, requiring chaining to generate suitable profiles. We see nonconvergence in many of the steps in the maximum likelihood estimation and the uncertainty quantification procedures, which computationally is flagged by longer runtimes. The successive chaining of estimates in the profile likelihood approach is particularly sensitive to this problem, since the result of the previous profile point (whether converged or not) is used as the initial iterate for the estimation of the next profile point. We have tried using restarts from surrounding iterates, and see that in some cases this can mitigate the problem, but this is not guaranteed. These convergence problems are exacerbated if the problem is relatively large, and we find that the explicit representation approach we use to save computational time has a large memory burden. We found that an RTO-style approach \citep{bardsley2014}, where the Jacobian must be used in the log-likelihood when resampling for a bootstrap, is infeasible on a workstation with 16 GB of RAM. 

\vspace{1em}
\textbf{Acknowledgements.} We wish to acknowledge Sapeer Mayron from the Samoa Observer for providing access to press releases that were not readily available. OJM would like to thank Elvar Bjarkason and Ruanui Nicholson for helpful discussions of the RTO and RML methods in the context of inverse problems.

\FloatBarrier

\bibliography{biblio}

\end{document}


\newcommand{\var}{\text{var}}
\newcommand{\Rz}{\mathcal{R}_{c}}
\newcommand{\tabbox}[2]{\vtop{\hbox{\strut #1}\hbox{\strut #2}}}

\begin{frontmatter}
\title{Supplementary Information}
\author[1]{David Wu \corref{cor}} 
\ead{dwu402@aucklanduni.ac.nz}
\author[2]{Helen Petousis-Harris}
\author[2]{Janine Paynter}
\author[1,3]{Vinod Suresh}
\author[1]{Oliver J. Maclaren}

\affiliation[1]{
    organization={Department of Engineering Science, University of Auckland},
    addressline={Grafton},
    city={Auckland},
    postcode={1010},
    country={New Zealand},
}
\affiliation[2]{
    organization={Department of General Practice and Primary Health Care, University of Auckland},
    addressline={Grafton},
    city={Auckland},
    postcode={1023},
    country={New Zealand},
}
\affiliation[3]{
    organization={Auckland Bioengineering Institute, University of Auckland},
    addressline={Grafton},
    city={Auckland},
    postcode={1010},
    country={New Zealand}
}

\cortext[cor]{Corresponding author}
\end{frontmatter}

\section{Application to SIR model}\label{sec:sir}

We can apply the method to a synthetic dataset generated from an SIR model. We generate data $y$ according to the model in \Cref{eq:sir}.

\begin{subequations}
    \begin{align}
        y &= \begin{pmatrix}
            S\\R
        \end{pmatrix} + \epsilon\\
        \frac{dx}{dt} = \frac{d}{dt}\begin{pmatrix}
            S\\I\\R
        \end{pmatrix} &= \begin{pmatrix}
            -\beta SI/N\\
            \beta SI/N - \alpha I\\
            \alpha I
        \end{pmatrix}\\
        \epsilon &\sim \mathcal{N}(0, \sigma^2 I)\\
        x(0) &= (N-1, 1, 0)^T
    \end{align}
    \label{eq:sir}
\end{subequations}

We note that the data generated has used a Gaussian error term for simplicity, although it does produce non-physical data, such as having negative values and decreasing total case trajectories.

The differential equation model of \Cref{eq:sir} is also used as our model for inference, with unknown $\theta = (\beta, \alpha), N,$ and $\sigma$, so the model is not misspecified in this case. We assume iid observation noise, and iid model error, giving the objective negative log-likelihood

\begin{equation}
    l(\theta, x) = \lVert w_0 (y - g(x)) \rVert^2 + \left\lVert w_1 \left(\mathcal{D}x - f(x, \theta)\right)\right\rVert^2,
\end{equation}
for the observation function $g(x) = \begin{pmatrix} S\\R \end{pmatrix}$, the derivative operator $\mathcal{D} \cdot = \frac{d}{dt}{\cdot}$, and some weights $w_0, w_1$ that correspond to the inverse standard deviation of the observation noise and model (discrepancy) error respectively. We note that in this model $\Rz = \frac{\beta}{\alpha N} = \mathcal{R}_0$, since there is no vaccination.

\begin{figure}[h]
    \centering
    \includegraphics[width=\textwidth]{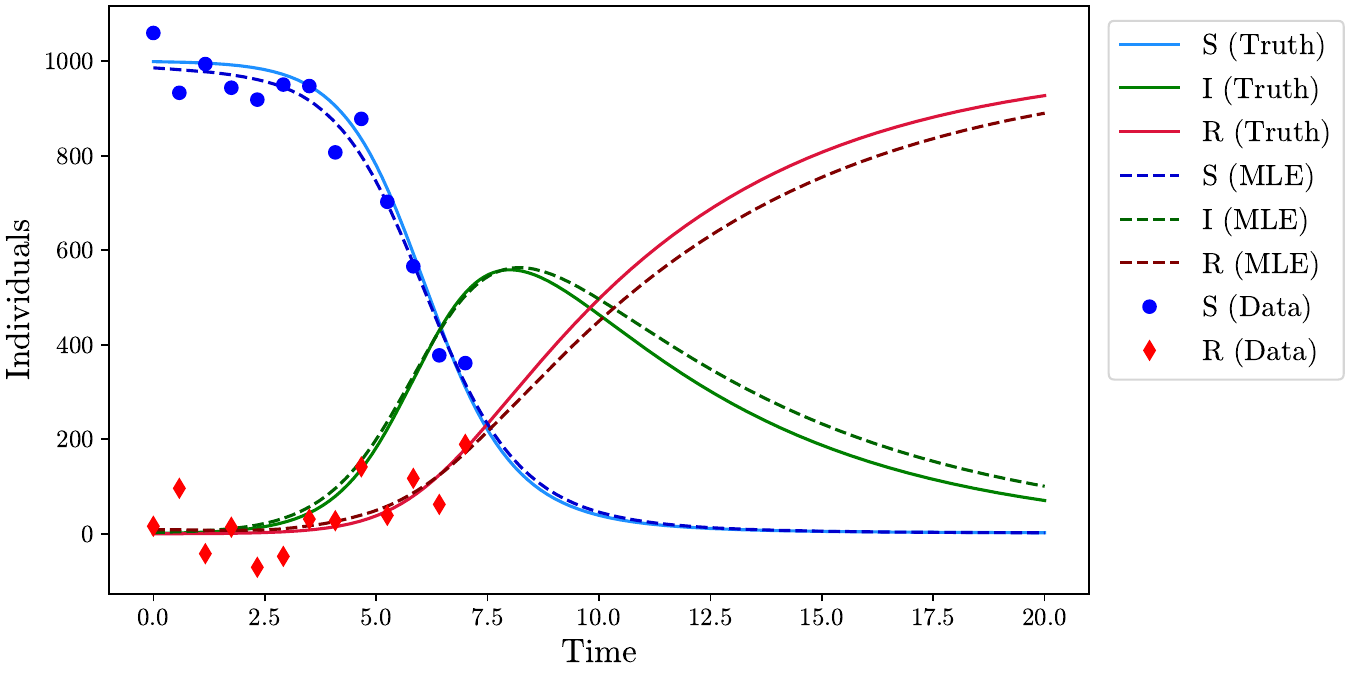}
    \caption{Fit of the SIR model as taken at three iteration of IRLS, plotted against the data used for fitting and the underlying (non-noisy) truth.}
    \label{fig:sir_fit}
\end{figure}

\begin{figure}[h]
    \centering
    \includegraphics[width=\textwidth]{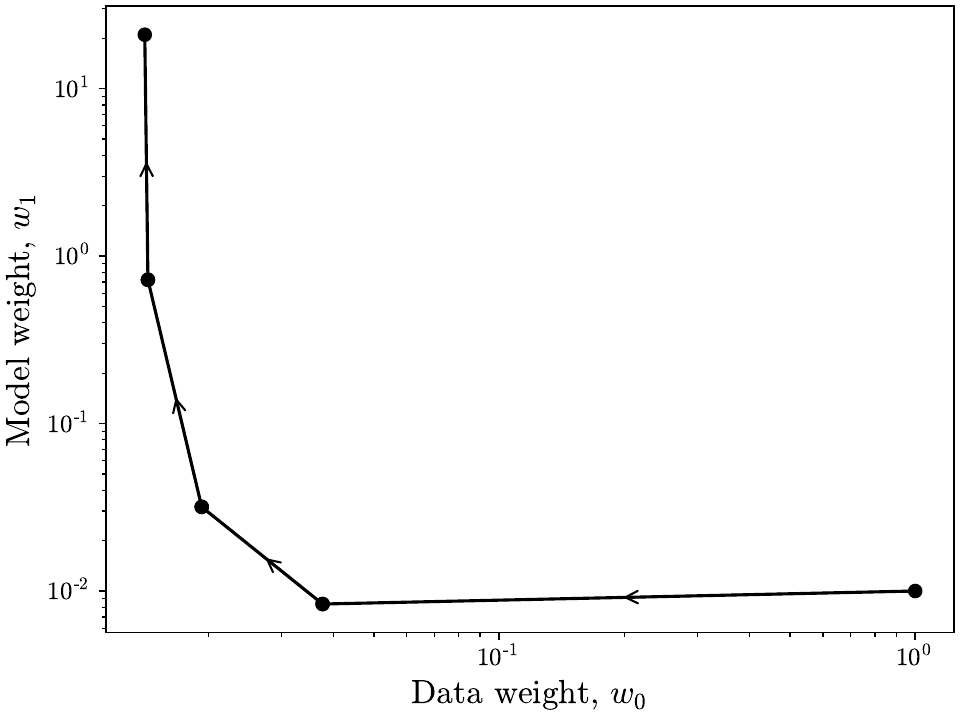}
    \caption{Log-log plot of weights $w_0, w_1$ of the SIR model log-likelihood as estimated at each iteration.}
    \label{fig:sir_weights}
\end{figure}

We generate data with $x(0) = (999, 1, 0), \alpha = 0.2, \beta = 1.3, \sigma = 50$.
We then fit this using initial iterates $x^{(0)}(t) = (1000,1000,1000), \alpha^{(0)} = \beta^{(0)} = 1$, initial weights $w_0=1, w_1=1\times10^{-2}$, and recover state estimates after 3 iterations plotted in \Cref{fig:sir_fit}. We provide a plot of the estimated weights at each iteration in \Cref{fig:sir_weights}. The weights lie on a L-shaped curve that is analogous to the L-curve in inverse problems literature \citep{hansen1992,hansen1993}.

\begin{figure}[h]
    \centering
    \includegraphics[width=\textwidth]{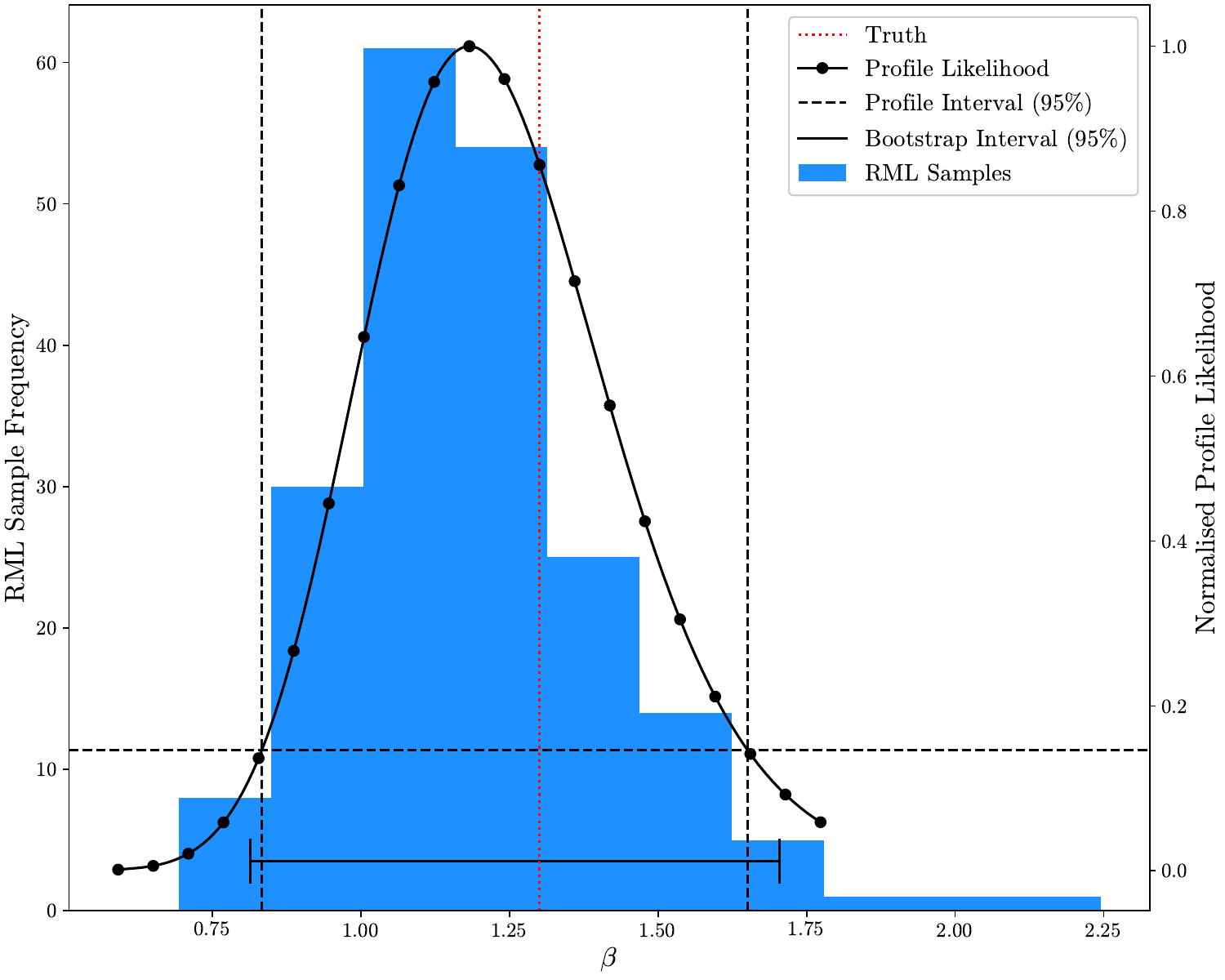}
    \caption{Profile likelihood and RML bootstrap samples $(n=200)$ of $\beta$ for the SIR model, with corresponding 95\% intervals}
    \label{fig:sir_beta_profile}
\end{figure}

\begin{figure}[h]
    \centering
    \includegraphics[width=\textwidth]{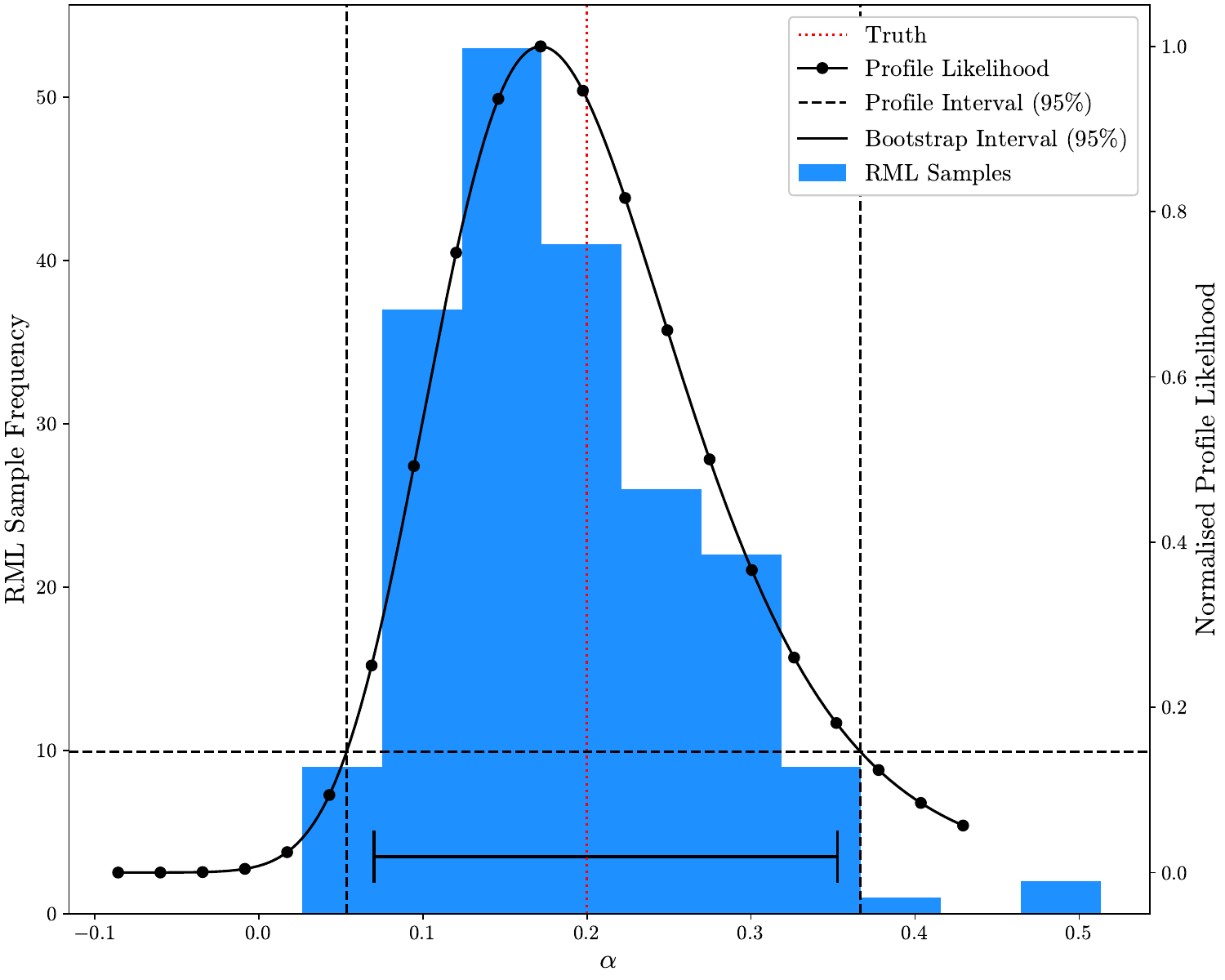}
    \caption{Profile likelihood and RML bootstrap samples $(n=200)$ of $\alpha$ for the SIR model, with corresponding 95\% intervals}
    \label{fig:sir_alpha_profile}
\end{figure}

\begin{figure}[h]
    \centering
    \includegraphics[width=\textwidth]{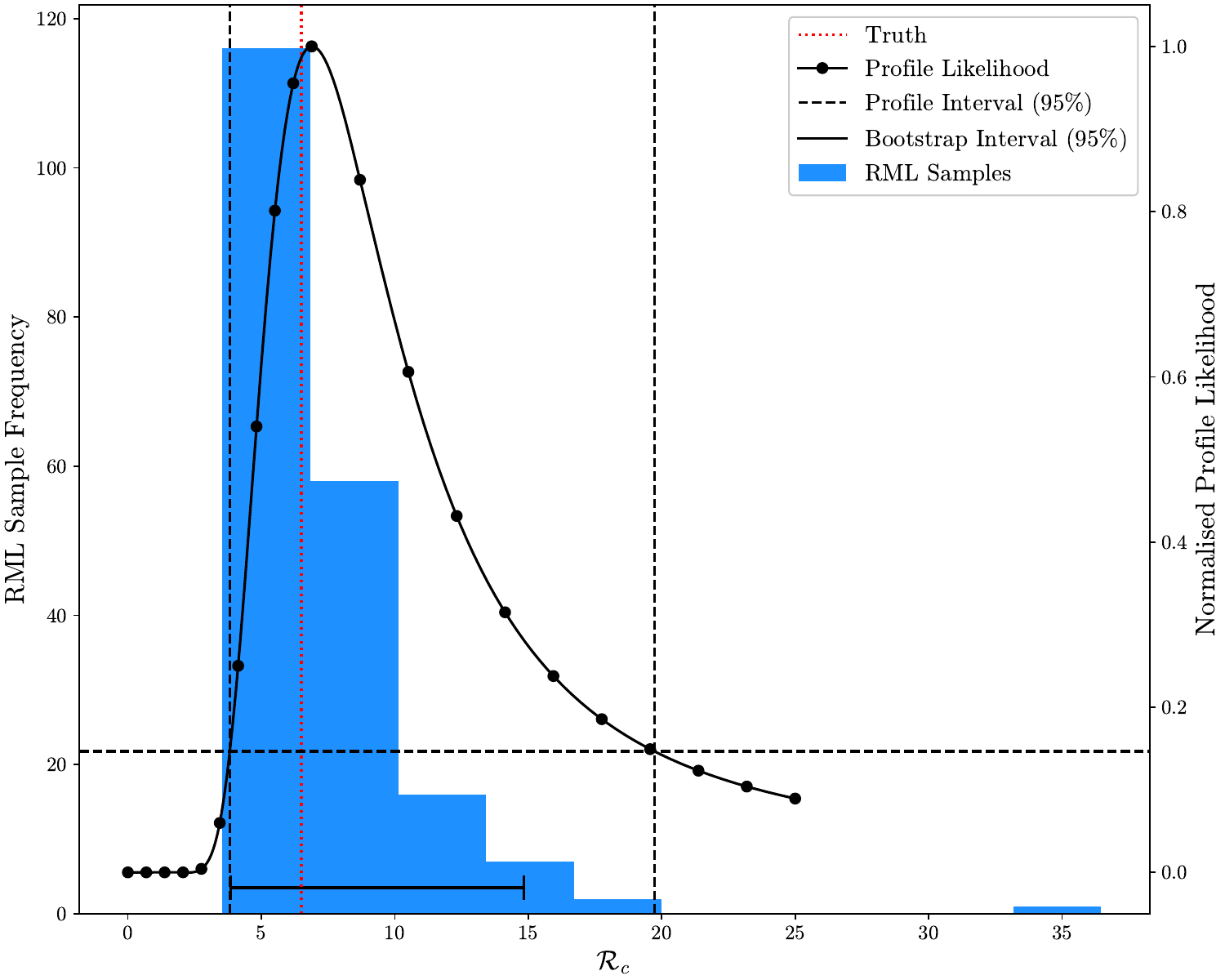}
    \caption{Profile likelihood and RML bootstrap samples $(n=200)$ of $\Rz$ for the SIR model, with corresponding 95\% intervals}
    \label{fig:sir_r0_profile}
\end{figure}

We then profile over the parameters $\beta$ and $\alpha$, as well as $\Rz$ in \Cref{fig:sir_beta_profile,fig:sir_alpha_profile,fig:sir_r0_profile}. We can also generate predictive uncertainties with RML bootstrapping, and also generate a bootstrap version of the parameter uncertainty. 95\% intervals derived from both the profile likelihood and the bootstrap samples are listed in \Cref{table:sir_intervals}.

\begin{table}[h]
    \centering
    \caption{Confidence intervals for the parameters of the SIR model as derived from profile likelihood and RML bootstrap samples $(n=200)$}
    \begin{tabular}{c c c c}
        \hline
        Value & Truth & Profile Likelihood 95\% Interval & Bootstrap 95\% Interval\\
        \hline
        $\beta$ & 1.3 & [0.8332, 1.650] & [0.8137, 1.704]\\
        $\alpha$ & 0.2 & [0.05349, 0.3666] & [0.07015, 0.3528]\\
        $\Rz$ & 6.5 &[3.819, 19.73] & [3.848, 14.84]\\
        \hline
    \end{tabular}
    \label{table:sir_intervals}
\end{table}

\begin{figure}[h]
    \centering
    \includegraphics[width=\textwidth]{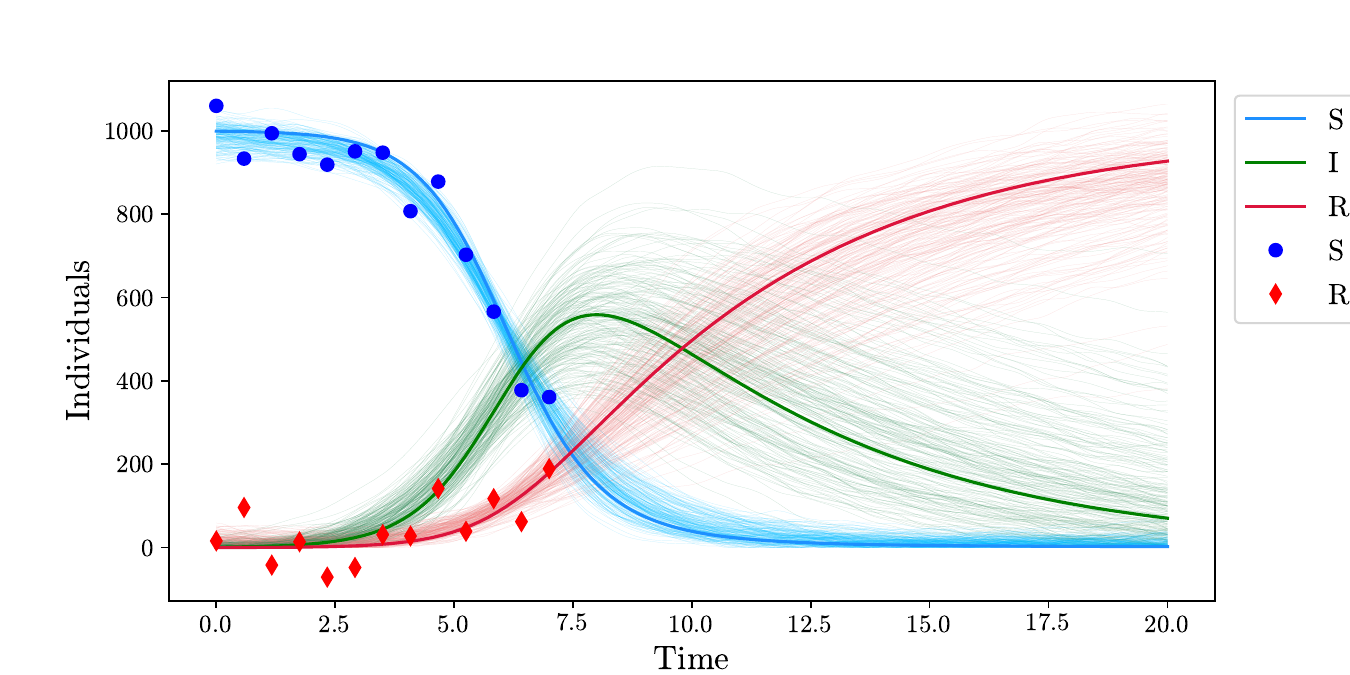}
    \caption{State estimates of the RML bootstrap samples for the SIR model ($n=200$)}
    \label{fig:sir_rml_samples}
\end{figure}

\begin{figure}[h]
    \centering
    \includegraphics[width=\textwidth]{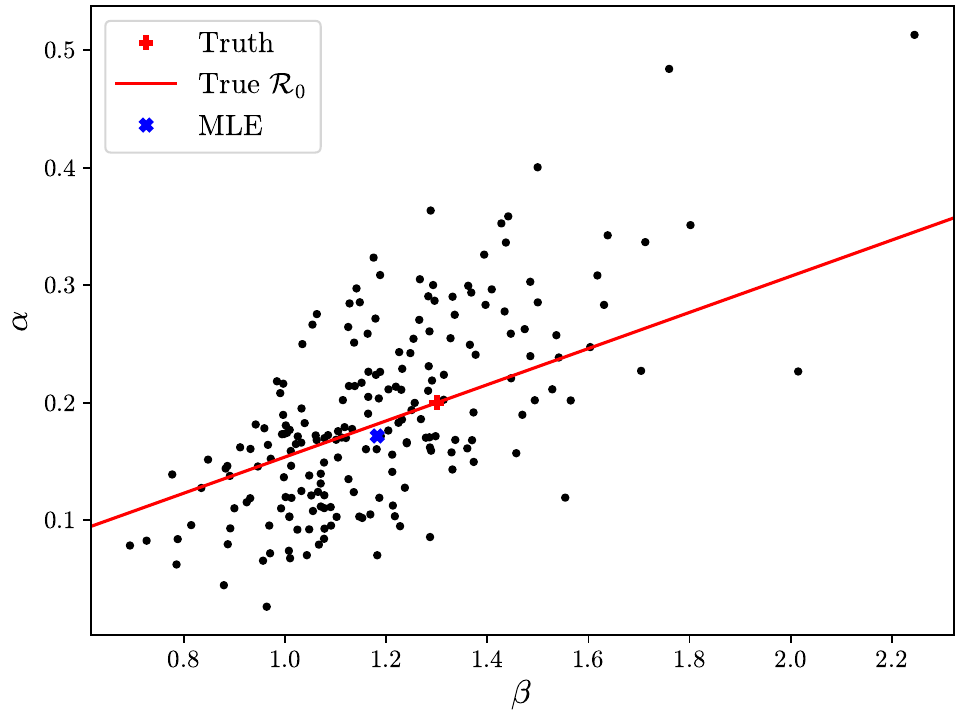}
    \caption{Parameters estimates of the RML bootstrap samples for the SIR model ($n=200$), with subset of parameters that correspond to the true $\mathcal{R}_0$ highlighted.}
    \label{fig:sir_rml_params}
\end{figure}

We note that the solve time for the MLE is 3 seconds, profiling takes a total of 1.5 seconds, and RML bootstrap sampling takes 37 seconds.

\FloatBarrier

\section{Additional Results and Figures of the Samoa Case Study}

Here we present some plots of validation and auxiliary information for the analysis in the main paper.

\begin{figure}[h]
    \centering
    \includegraphics[width=\textwidth]{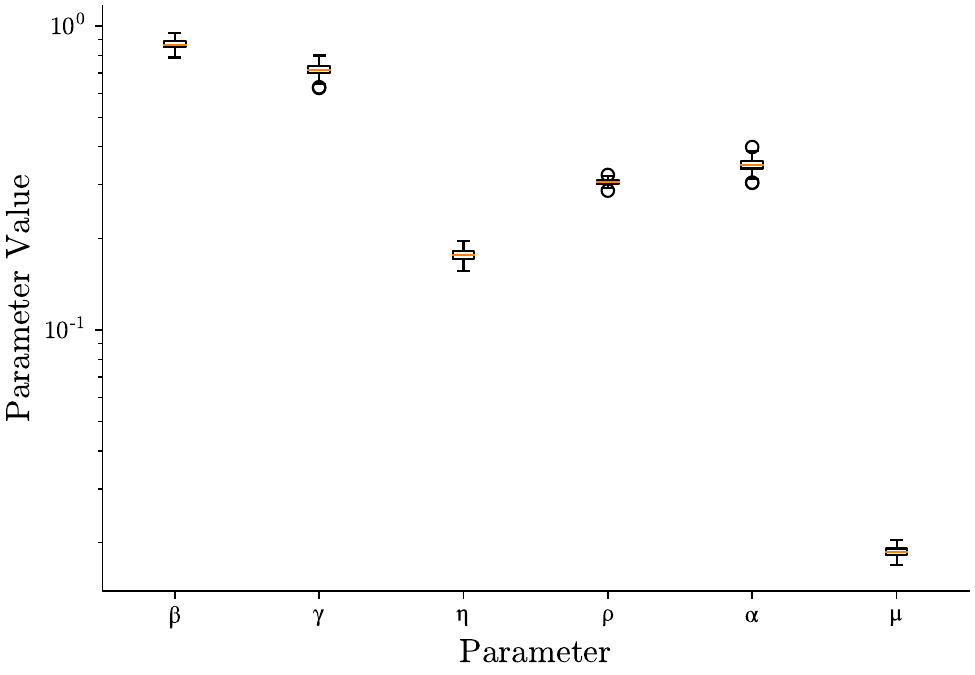}
    \caption{RML samples of the parameter estimates for the Samoa data}
    \label{fig:samoa_parameters_rml}
\end{figure}

\begin{figure}[h]
    \centering
    \includegraphics[width=\textwidth]{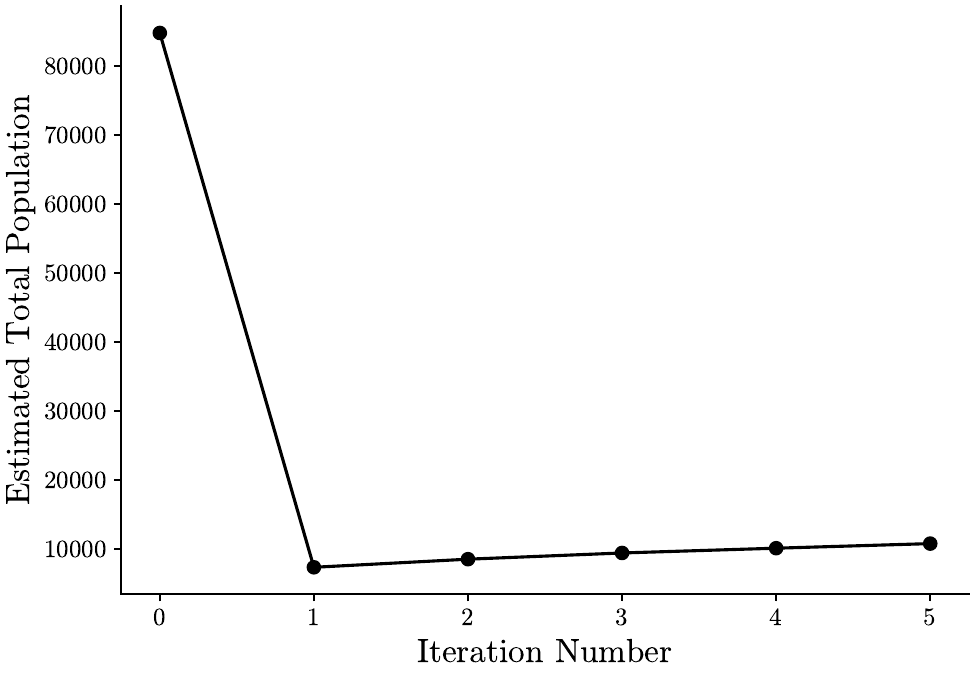}
    \caption{Estimated total population along iterations of IRLS for estimating MLE for Samoa data}
    \label{fig:est_totpop_apdx}
\end{figure}

\begin{figure}[h]
    \centering
    \includegraphics[width=\textwidth]{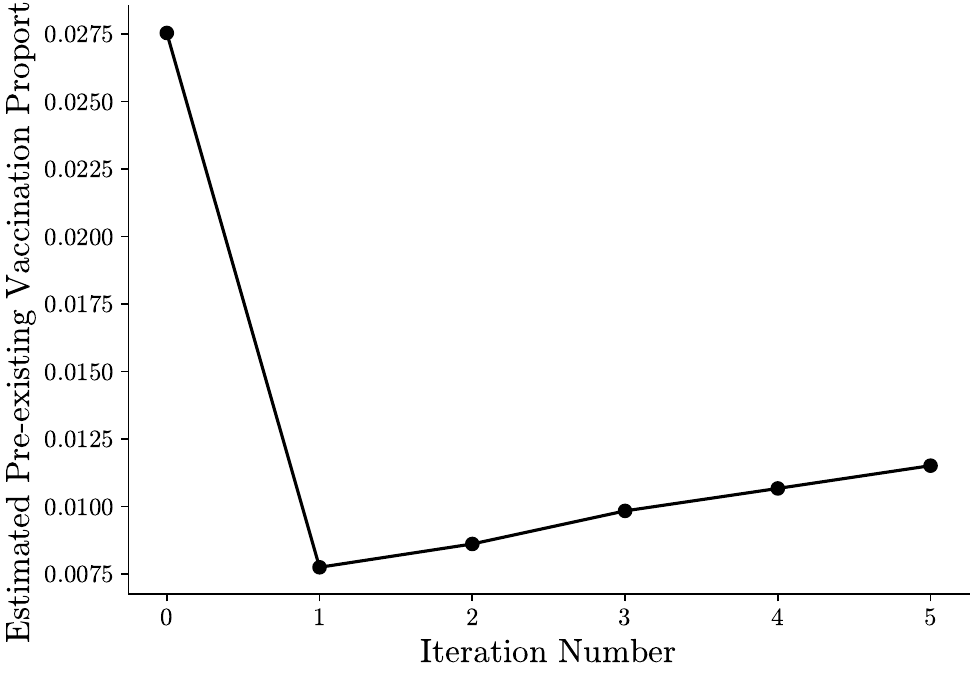}
    \caption{Estimated pre-existing vaccination proportion along iterations of IRLS for estimating MLE for Samoa data}
    \label{fig:est_vacc_apdx}
\end{figure}

\begin{figure}[h]
    \centering
    \includegraphics[width=\textwidth]{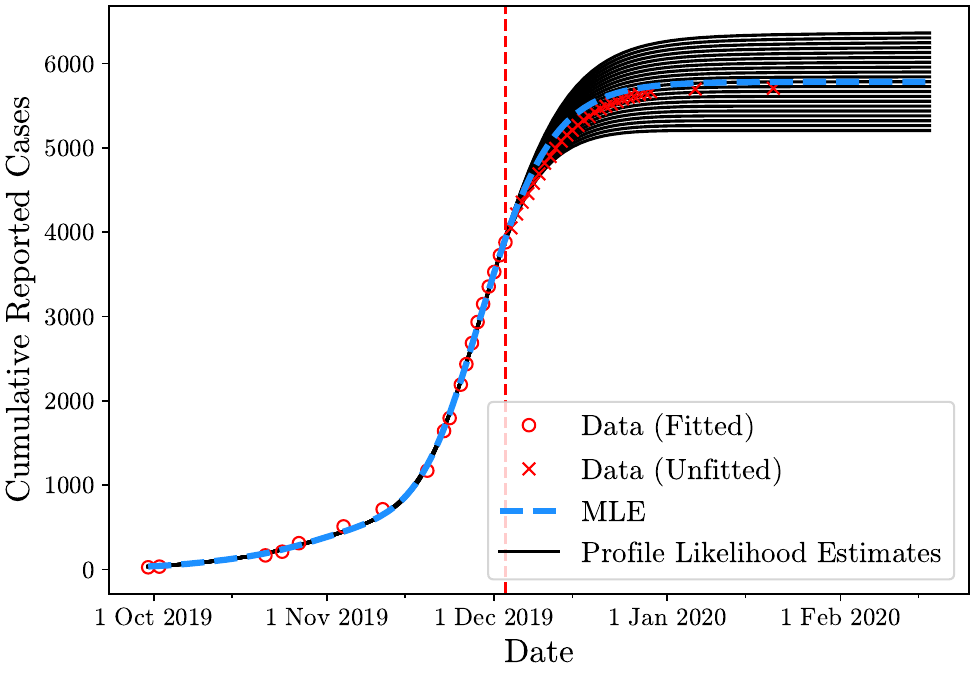}
    \caption{Estimated trajectories for cumulative cases over time along the profile of total cases}
    \label{fig:tcprofile_esttraj}
\end{figure}

\begin{figure}[h]
    \centering
    \includegraphics[width=\textwidth]{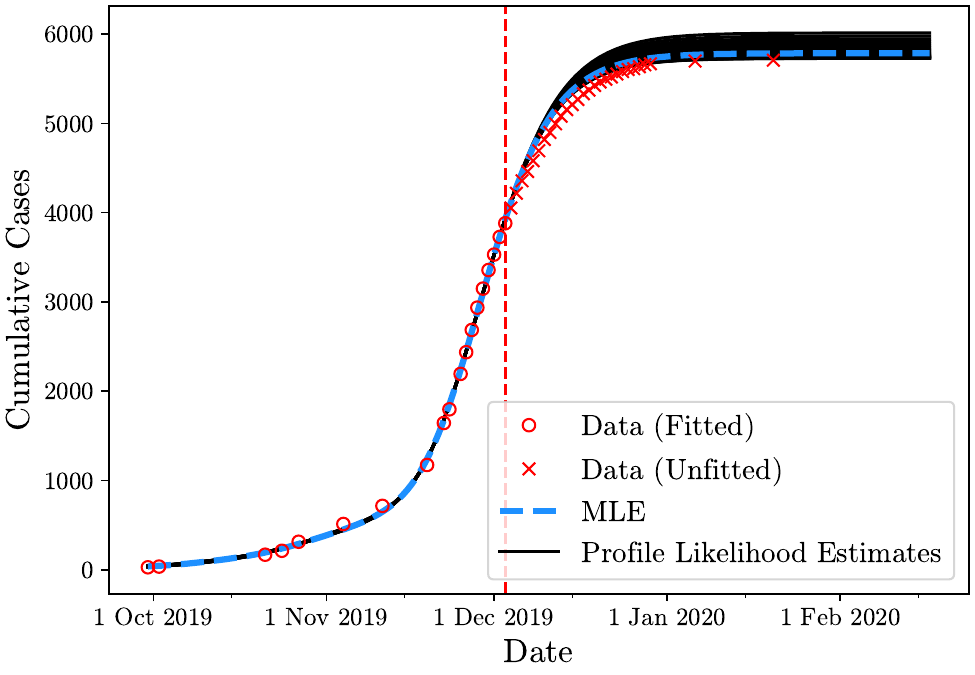}
    \caption{Estimated trajectories for cumulative cases over time along the profile of $\Rz$}
    \label{fig:r0profile_esttraj}
\end{figure}

\FloatBarrier

\section{Results from Synthetic Study (Regularisation)}

We use the ODE model used to fit the Samoa data, and generate synthetic data with parameters and initial conditions listed in \Cref{table:synthetic_samoa_parameters}.

\begin{table}[h]
    \caption{Parameters and Initial Conditions for the Synthetic Study of the Samoa Measles Model}
    \label{table:synthetic_samoa_parameters}
    \centering
    \begin{tabular}{|c|c|}
        \hline
        Parameter/Initial Condition & Value \\
        \hline
        $\beta$ & 5 \\
        $\gamma$ & 0.25 \\
        $\eta$ & 0.05 \\
        $\delta$ & 0.2 \\
        $\alpha$ & 0.2 \\
        $\mu$ & 0.015 \\
        \hline
        $S(0)$ & 20 000\\
        $E(0)$ & 5 \\
        $I(0)$ & 5 \\
        $R(0)$ & 100 000\\
        $H(0)$ & 0 \\
        $G(0)$ & 0 \\
        $D(0)$ & 0 \\
        $I_c(0)$ & 0 \\
        $H_c(0)$ & 0 \\
        \hline
    \end{tabular}
\end{table}

We add noise by using a Poisson noise model, and only provide the data in the first third of the time domain for fitting. We also partially observe the data in the same way as the actual data ($S, E, I, R$ states not observed).
\begin{equation}
    y^* \sim Poisson(y)
\end{equation}

We fit the model with an additional regularisation term for the parameter $\gamma$:
\begin{equation}
    -2\log\mathcal{L}(x, \theta) = \lVert L_d(y^* - g(x)) \rVert^2 + \lVert L_m(\mathcal{D}x - f(x, \theta)) \rVert^2 + \lVert \rho (\gamma - 0) \rVert^2
\end{equation}

where $g$ is the function representing the partial observation of the states $x$, $f$ is the ODE model, and $\rho$ is the regularisation strength. As usual, $x$ is projected on a spline basis $\Phi$ such that $x = \Phi c$.

We compute the profiles over $\gamma$ as $\rho$ is increased from 0 to 24. We see that when $\rho = 0$, even with appropriate bounds on $L_d$ and $L_w$, the profile does not capture the true value of $\gamma$. Appropriate values of $\rho$ lie in the range of 15 to 20, as we see the validation error increasing greatly above this interval.

In the case study, we penalise both $\beta$ and $\gamma$ using the log-likelihood:
\begin{equation}
    -2\log\mathcal{L}(x, \theta) = \lVert L_d(y^* - g(x)) \rVert^2 + \lVert L_m(\mathcal{D}x - f(x, \theta)) \rVert^2 + \lVert \rho (\gamma - 0) \rVert^2 + \lVert \rho (\beta - 0) \rVert^2.
\end{equation}

\begin{figure}[h]
    \centering
    \begin{subfigure}{0.45\textwidth}
        \includegraphics[width=\textwidth]{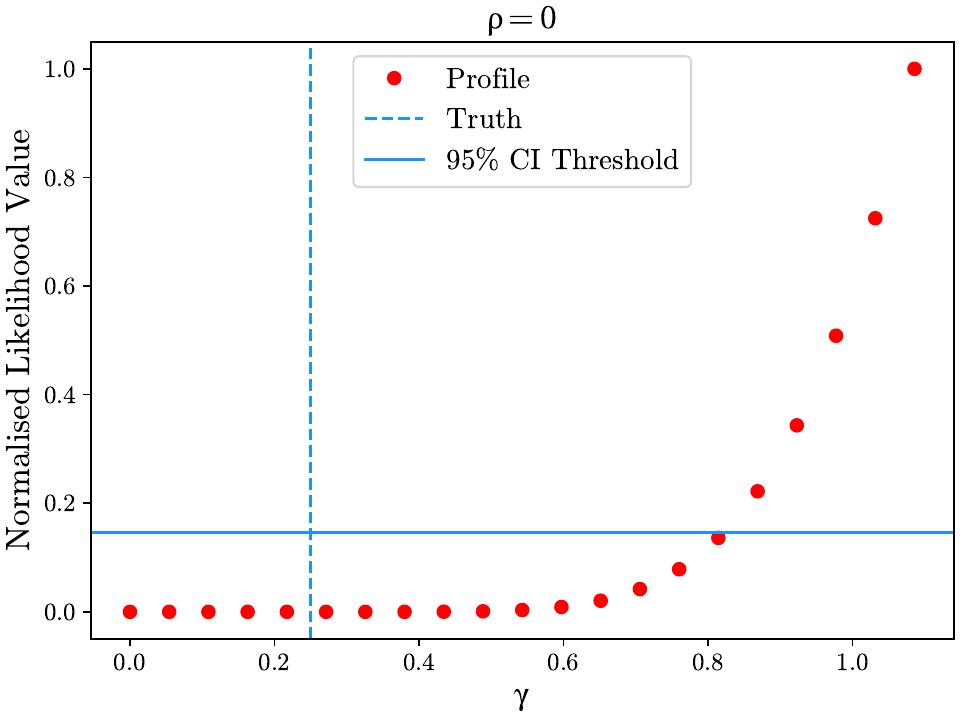}
    \end{subfigure}
    \begin{subfigure}{0.45\textwidth}
        \includegraphics[width=\textwidth]{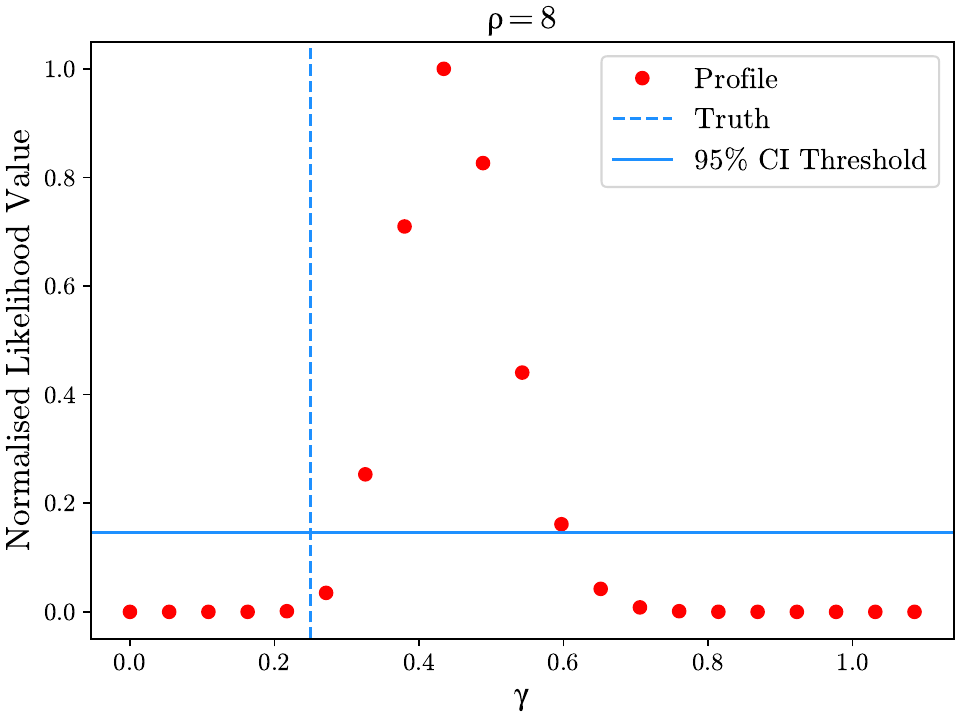}
    \end{subfigure}
    \begin{subfigure}{0.45\textwidth}
        \includegraphics[width=\textwidth]{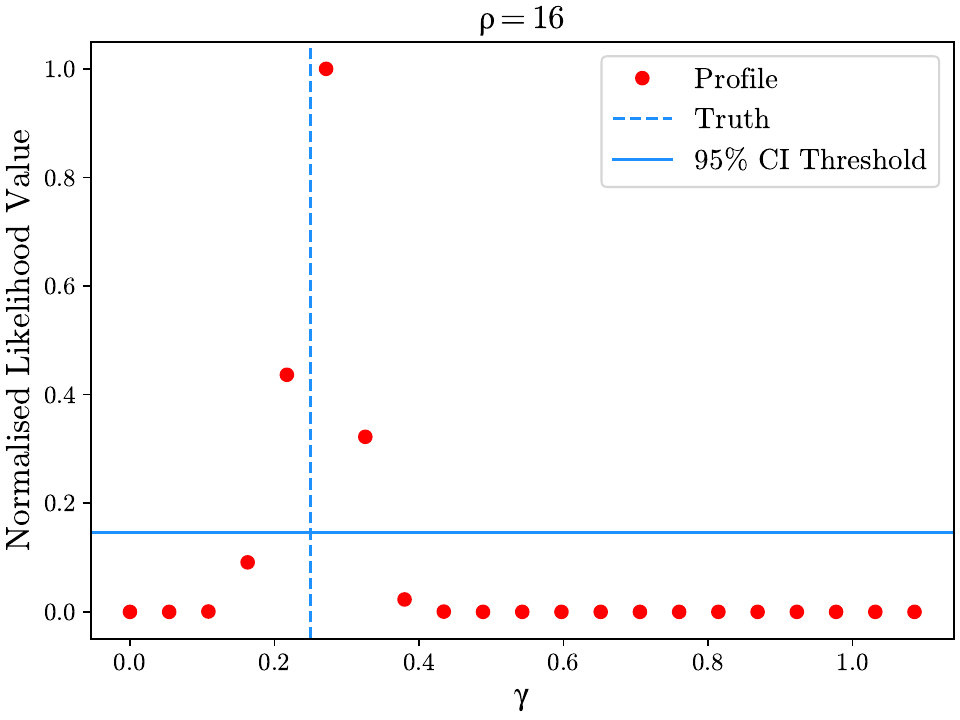}
    \end{subfigure}
    \begin{subfigure}{0.45\textwidth}
        \includegraphics[width=\textwidth]{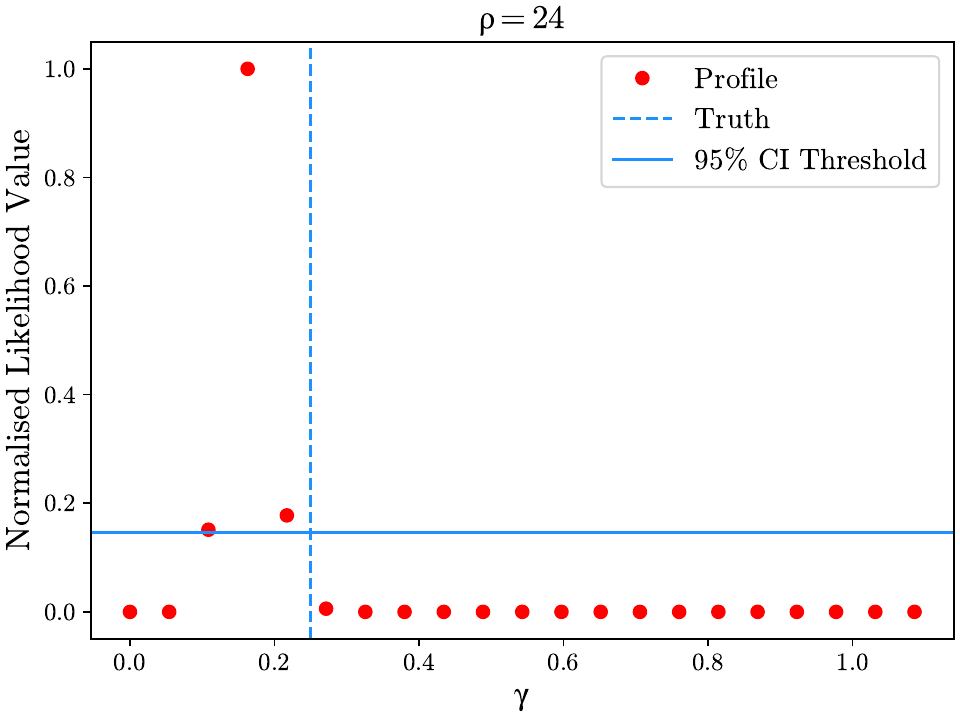}
    \end{subfigure}
    \caption{Profiles of $\gamma$ with regularisation strengths $\rho = \{0, 8, 16, 24\}$.}
\end{figure}

\begin{figure}[h]
    \centering
    \includegraphics[width=0.8\textwidth]{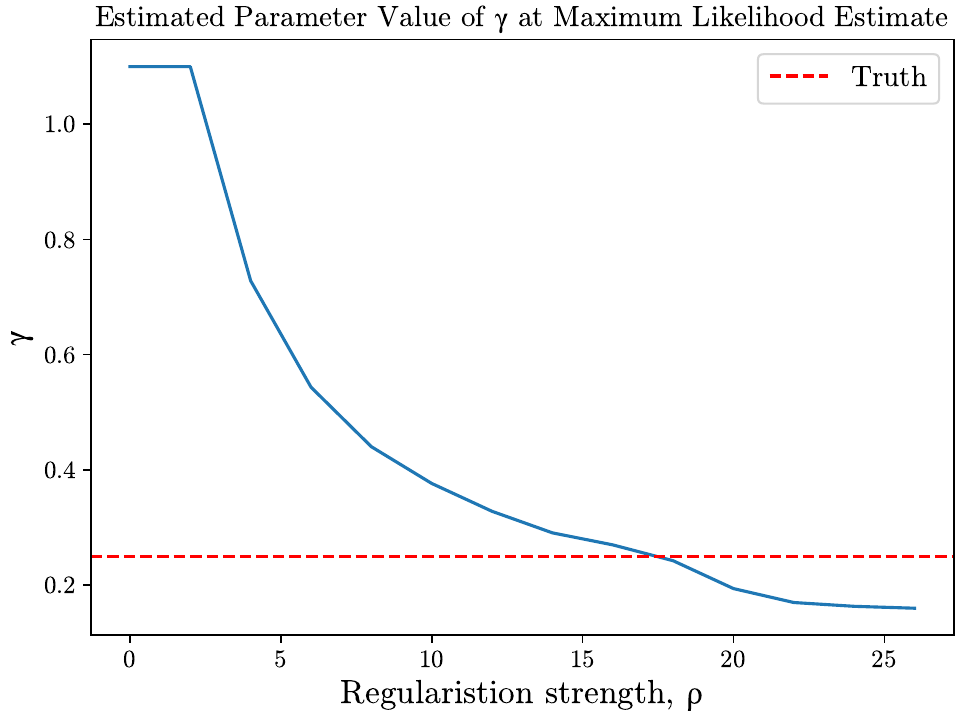}
    \caption{Estimate of $\gamma$ at the maximum (normalised) likelihood value over the range of $\gamma$ profiled as regularisation strength $\rho$ is varied.}
\end{figure}

\begin{figure}[h]
    \centering
    \includegraphics[width=0.8\textwidth]{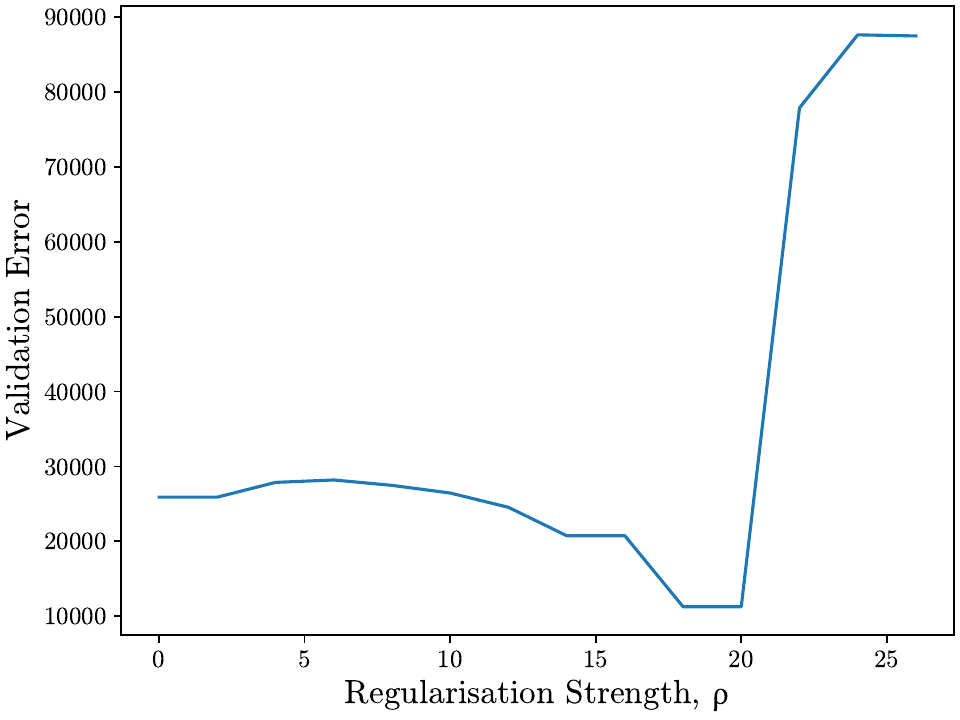}
    \caption{Validation error as the regularisation strength $\rho$ is varied. Validation error is computed as $\lVert y - x \rVert_1$ over the entire time domain.}
\end{figure}

\FloatBarrier

\section{Observation Window and Identifiability}\label{sec:obsv_window}
It is known that the predictive uncertainty of epidemics shrinks drastically as we pass the ``turning point'' of the epidemic curve \citep{wilke_bergstrom_2020}. We perform an analysis comparing the predictive error of the maximum likelihood estimator as we change the observation window.

Data is provided as daily observations at $t \in [0, T]$, where $T \in [10, 11, 12, ..., 100]$, and observed through a Poisson error model, where the underlying truth is generated over the time interval $[0, 100]$, with initial conditions $x_0$ and parameters $p_{true}$. The data is then perturbed and refit using the RML-bootstrap method.

\begin{subequations}
\begin{align}
    y &\sim Poisson(x_{obsv}), \\
    x_{obsv} &= \{ S, R \},\\
    \frac{dS}{dt} &= -\beta S\frac{I}{N},\\
    \frac{dE}{dt} &= \beta S\frac{I}{N} - \gamma E,\\
    \frac{dI}{dt} &= \gamma E - \alpha I,\\
    \frac{dR}{dt} &= \alpha I,
\end{align}
\end{subequations}
where $x = \{S, E, I, R\}$ and $p = \{\beta, \gamma, \alpha\}$. 

\begin{figure}[h]
    \centering
    \includegraphics[width=\textwidth]{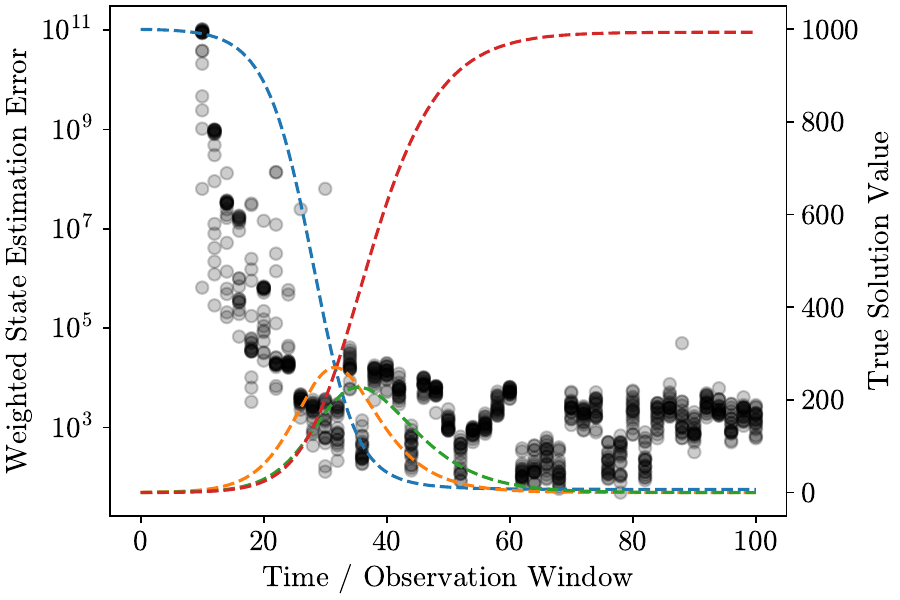}
    \caption{Weighted state estimation error from 20 RML bootstrap samples overlaid on true trajectory}
    \label{fig:sest_rml}
\end{figure}

\begin{figure}[h]
    \centering
    \includegraphics[width=\textwidth]{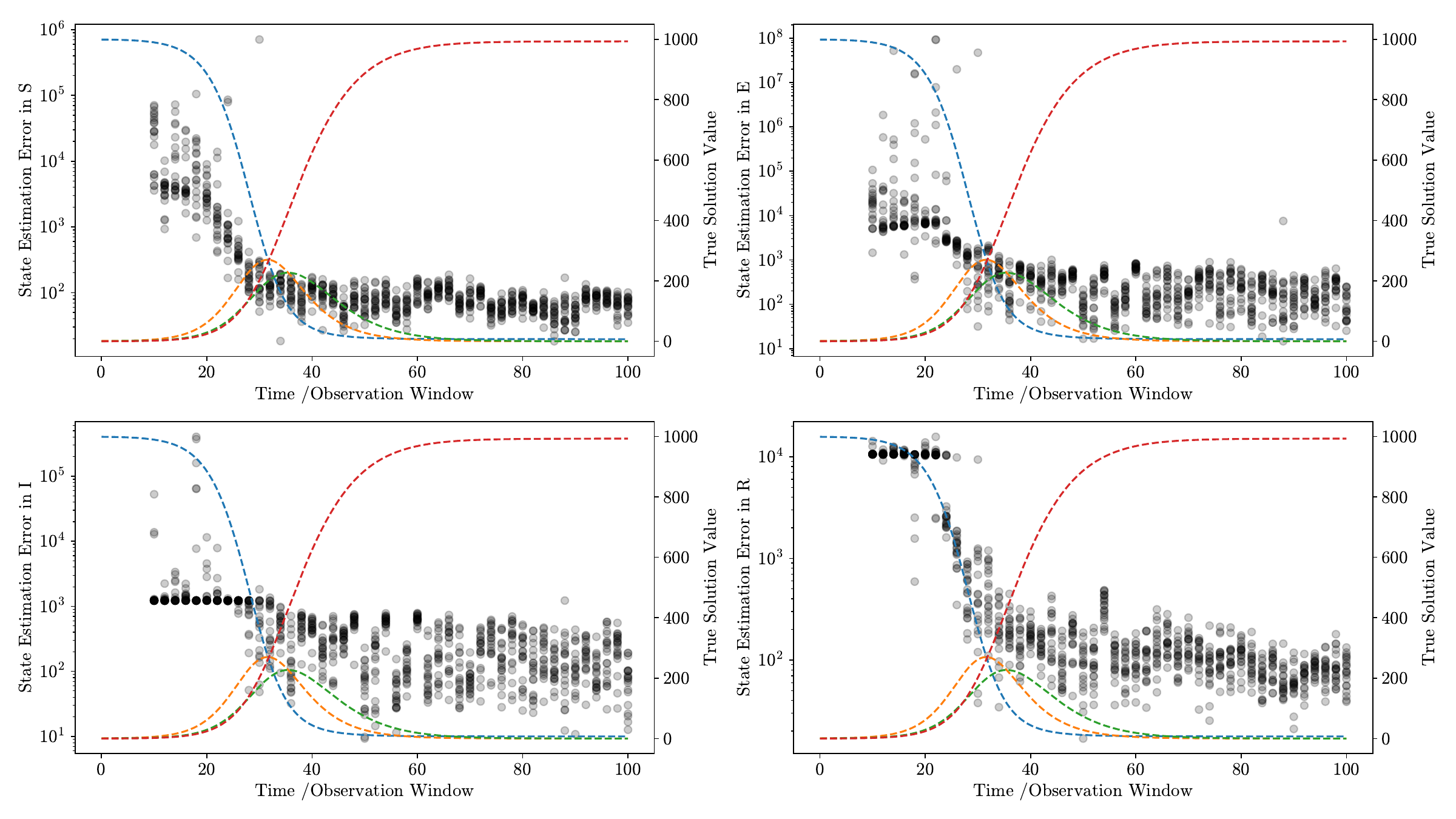}
    \caption{State estimation of each of the states from 20 RML bootstrap samples overlaid on true trajectory. (top-left) S, (top-right) E, (bottom-left) I, (bottom-right) R.}
    \label{fig:s4est_rml}
\end{figure}

The predictive error $e_x(T)$ is computed from the MLE $x_{MLE, T}$ for a given observation window $[0, T]$ as

\begin{equation}
    e_x(T) = \lVert x(t) - x_{MLE, T}(t) \rVert^2_{\Gamma^{-1}}
\end{equation}

where $t \in [0, 100]$. This is presented in \Cref{fig:sest_rml}, with individual states' errors in \Cref{fig:s4est_rml}.

We also compute the error $e_p(T)$ in the parameter estimates from the MLE $p_{MLE, T}$ as 

\begin{equation}
    e_p(T) = \lVert p_{true} - p_{MLE, T} \rVert^2.
\end{equation}

\begin{figure}[h]
    \centering
    \includegraphics[width=\textwidth]{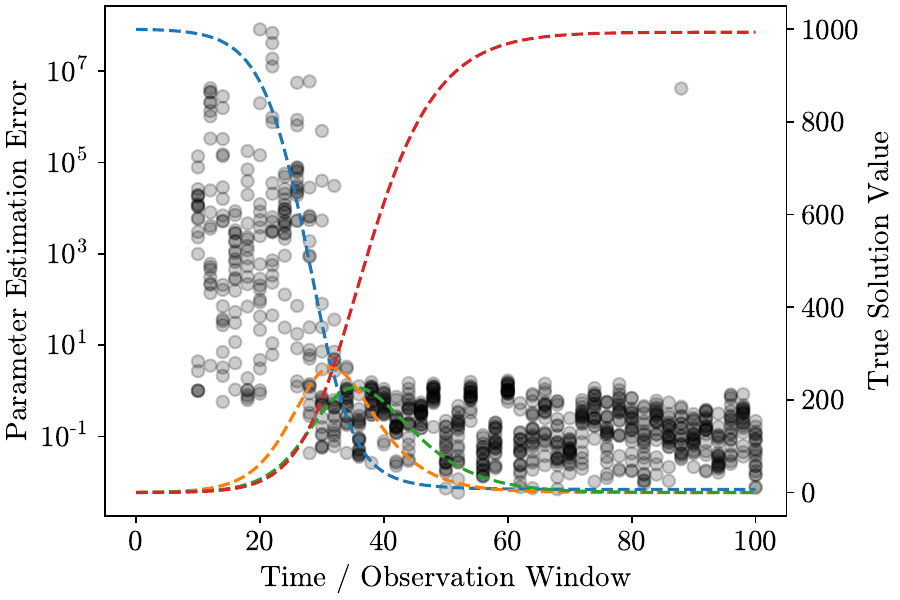}
    \caption{Parameter estimation error from 20 RML bootstrap samples overlaid on true trajectory}
    \label{fig:pest_rml}
\end{figure}

This is presented in \Cref{fig:pest_rml}

We see that the state estimation error seems to decrease smoothly with an increasing observation window, but there seems to be a distinct difference in the parameter estimation errors past the turning point of the epidemic curve.

\FloatBarrier

\section{Implementation Detail}\label{sec:implement}
Implementation of this method can be found on GitHub at \url{https://github.com/dwu402/pypei}.
The Samoa case study is presented at \url{https://github.com/dwu402/samoa-books}.

\subsection*{Basis and Objective Function Representation}
The problem is represented in \texttt{CasADi} \cite{andersson2019} by computing $\Phi$ over a specified time grid $\mathbb{T} = \{t_1, t_2, \dots, t_N\}$, by evaluating the generated (\texttt{MX}) spline basis $\mathbb{S} = \{\phi_1, \phi_2, \dots, \phi_K \}$ over $\mathbb{T}$. This has the computational advantage of reducing the calls needed when computing the objective function. The derivative basis $D\Phi$ was computed similarly by evaluating the derivatives of $\mathbb{S}$ over $\mathbb{T}$. $c$ and $\theta$ were represented as \texttt{SX} objects.

The basis is constructed with uniformly spaced knots. In the case study in the main text, as well the results in the supplementary material, there are $K=40$knots and $\mathbb{T}$ is constructed with $N=200$. As stated in the main text, we set $t_N$ as 1.25 of the final data point in the Samoan case study, that is extrapolating past the end of the epidemic. The expectation is that this is sufficiently far in the future that the underlying ODE model would constrain it towards a stable equilibrium, and although this is not strictly true, in practice it is very often the case. In the synthetic studies, $t_N$ is set equal to the final time point of the synthetic data, as this is sufficient for validation.

The objective object was constructed by constructing further \texttt{SX} objects: \texttt{y0} representing data, \texttt{W} representing the whitening matrices. The objective is then constructed as

\begin{center}
    \texttt{sum}(\texttt{sumsqr}(\texttt{W} @ (\texttt{y0} - \texttt{y})))
\end{center}

where @ is the matrix multiplication operator, \texttt{y} represents the state- and parameter-based components. In our case, \texttt{y} is the array [$g(c)$, $D\Phi c - f(\Phi c, \theta)$]. For our case study, we implement $g(c)$ as $\Psi c$, where $\Psi$ is constructed similarly to $\Phi$, by evaluating $\mathbb{S}$ over the time points $t_y$ of the data. $f$ is the ODE model. 

\subsection*{Covariance Structure}
We choose $W$ to be the Cholesky whitening matrices so that $LL^T = \Gamma^{-1}, MM^T = \Sigma^{-1}$, and $L$ and $M$ are triangular. This allows for $\log|\Gamma|^{-1/2} = \log|L|$ to be computed efficiently as the sum of the log of the diagonal elements of $L$.
For the Samoan measles case study, $\Gamma$ and $\Sigma$ are constructed to be diagonal, with some structure that allows for different states (of the ODE model) to have different variances. We specify elements of (the diagonal) $W$ so that states with similar sizes have the same model error. Specifically, we have $\var(S) = \var(I_c) = \var(R)$, $\var(E) = \var(I)$, and $\var(G) = \var(H_c)$. This is akin to the weights $w_i$ that are applied in \citet{hooker2011}.

\subsection*{MLE Detail}
We solve the problem iteratively with IRLS, using the \texttt{IPOPT} \cite{waechter2006} plugin to \texttt{CasADi} to solve the objective, and then reweighting $L$ and $M$ so that they match the estimated optimal covariances at each iteration. Initial iterates for the diagonal elements of $L$ and $M$ were set at $1$ and $2\times 10^{-2}$ respectively. Initial iterates of $c$ were set at $10000$, and $\theta$ at $0.1$. Iterations were repeated for $N = 5$ iterations. 

\subsection*{Profiling Detail}
Profiling was done by gridding the quantity of interest $\omega$ over the interval $[v \omega_{MLE}, (1+v) \omega_{MLE}]$ uniformly for some constant $v$. For our implementation, beacuse we are using \texttt{IPOPT}, we can do this by setting bounds or constraints on $\omega$ at each grid point. In practice, we grid in two subintervals $(\omega_{MLE} , v \omega_{MLE}]$ and $(\omega_{MLE} , (1+v)) \omega_{MLE}]$. This is so we can use the MLE as the initial iterate, and compute each subsequent profile point by using the solution of the previous profile point, for both numerical stability and computational efficiency. To compute the confidence intervals, the profiled points are first interpolated using a cubic spline implementation in Scipy, so that root-finding algorithms can be used. The ``normalised'' likelihoods $\mathcal{L}^*$, scaled by the maximum likelihood, are used for the computation of the confidence intervals:

\begin{equation}
    \mathcal{L}^*(\omega) = \frac{\mathcal{L}(\omega)}{\mathcal{L}(\omega_{MLE})} = \exp\left(-\frac{1}{2} (\mathcal{L}(\omega) - \mathcal{L}(\omega_{MLE}))\right)
\end{equation}

The confidence intervals can then be computed by solving

\begin{equation}
    \mathcal{L}^*(\omega) = \exp\left(-\frac{1}{2} \chi^2_{(q, (1-\alpha))})\right)
\end{equation}

where $q$ is the number of degrees of freedom ($1$), and $\alpha$ is the confidence level ($0.95$). For our values, this is $\chi^2_{(1, 1-0.95)} = 3.84$, or roughly $\mathcal{L}^*(\theta) = 0.15$.

For the bivariate (joint) profile, a similar procedure was undertaken, splitting the domain into four quadrants, and iteratively computing profile points moving away from the MLE estimate. The 95\% confidence region is taken at the same $\mathcal{L}^*(\theta) = 0.15$ level, and is automatically interpolated from the profiled grid points by the \texttt{contour} function in the \texttt{matplotlib} library \cite{hunter2007}.

\subsection*{RML Detail}
For the RML bootstrapping procedure, we use the Cholesky whitening matrices ($W$) to generate the samples of $e$ and $\nu$. This can be done simply by drawing $N$ samples from the standard normal distribution $Z \sim \mathcal{N}(0, I)$ and computing the samples $\tilde Y$ by solving the linear system $W\tilde Y = Z$. The samples are then fitted by minimising the objective function with the resampled data $\tilde Y$ from the initial iterate of the MLE.

We found that implementing an RTO \cite{bardsley2014} sampling technique was not possible on our machine (Intel i7-7700 CPU, 15.5 GiB memory) due to the dense $\bar Q$ matrix, which was too memory-intensive to algorithmically differentiate with the explicit \texttt{SX} representation in \texttt{CasADi}.

\FloatBarrier

\bibliography{biblio}